\documentclass[12pt]{article}
\usepackage[bookmarksnumbered,plainpages]{hyperref}
\usepackage{latexsym}
\usepackage{amsfonts}
\usepackage{amssymb}
\usepackage{amsbsy}
\usepackage{amsmath}   
\usepackage{supertabular}
\usepackage{theorem}
\renewcommand{\theequation}{\arabic{section}.\arabic{equation}}

\makeatletter
\@addtoreset{equation}{section}
\makeatother

\newtheorem{Def}{Definition}[section]

\def\p{\partial}
\def\dfrac#1#2{{\displaystyle\frac{#1}{#2}}}
\def\stTD#1#2{\hbox to 0em{\mathsurround=0em $\stackrel{#1}{\makebox[0pt]{} #2}$\hss} \phantom{#2}}\def\stscript#1#2{\hbox to 0em{\mathsurround=0em ${\scriptstyle\stackrel{#1}{\makebox[0pt]{} #2}}$\hss} \phantom{#2}}\def\stscriptscript#1#2{\hbox to 0em{\mathsurround=0em ${\scriptscriptstyle\stackrel{#1}{\makebox[0pt]{} #2}}$\hss} \phantom{#2}}\def\st#1#2{\mathchoice{\stTD{#1}{#2}}{\stTD{#1}{#2}}{\stscript{#1}{#2}}{\stscriptscript{#1}{#2}}}
\def\comb#1#2#3{{\mathsurround 0pt\hbox to 0pt {\hspace*{#3}\raisebox{#2}{${#1}$}\hss}}}
\def\combs#1#2#3{{\mathsurround 0pt\hbox to 0pt {\hspace*{#3}\raisebox{#2}{${\scriptstyle #1}$}\hss}}}
\def\combss#1#2#3{{\mathsurround 0pt\hbox to 0pt {\hspace*{#3}\raisebox{#2}{${\scriptscriptstyle #1}$}\hss}}}
\def\e#1{\mathrm{e}^{#1}}
\def\ii{\mathrm{i}}

\def\df{\mathrm{d}}

\def\metr{\mathfrak{m}}
\def\Fem{F}
\def\bFem{\mathbf{F}}
\def\fem{G}
\def\bfem{\mathbf{G}}
\def\bcdot{\mathchoice{\mathbin{\boldsymbol{\comb{\cdot}{0.22ex}{0.57ex}\Diamond}}}{\mathbin{\boldsymbol{\comb{\cdot}{0.22ex}{0.57ex}\Diamond}}}{\mathbin{\boldsymbol{\combs{\cdot}{0.13ex}{0.35ex}\Diamond}}}{}{}}
\def\bwedge{\mathchoice{\mathbin{\combs{\boldsymbol{\backslash}}{0.2ex}{0.1ex}\boldsymbol{\wedge}}}{\mathbin{\combs{\boldsymbol{\backslash}}{0.2ex}{0.1ex}\boldsymbol{\wedge}}}{\mathbin{\combss{\boldsymbol{\backslash}}{0.1ex}{0ex}\boldsymbol{\wedge}}}{}{}}
\def\Vol{\mathchoice{\combs{\square}{0.15ex}{0.2ex} {\rm V}}{\combs{\square}{0.15ex}{0.2ex} {\rm V}}{\combss{\square}{0.12ex}{0.095ex} {\rm V}}{}{}}

\def\dVol{{\rm d}\hspace{-0.3ex}\Vol}

\def\Eem{E}
\def\bEem{\mathbf{E}}
\def\Hem{H}
\def\bHem{\mathbf{H}}
\def\Dem{D}
\def\bDem{\mathbf{D}}
\def\Bem{B}
\def\bBem{\mathbf{B}}
\def\je{{\mathsurround 0pt\lower.0ex\hbox{${\scriptscriptstyle e}$}\mspace{-4.5mu}j}}
\def\bje{{\mathsurround 0pt\lower.0ex\hbox{${\scriptscriptstyle \mathbf{e}}$}\mspace{-3.4mu}\mathbf{j}}}
\def\jm{{\mathsurround 0pt\lower.0ex\hbox{${\scriptscriptstyle m}$}\mspace{-7.0mu}j}}
\def\bjm{{\mathsurround 0pt\lower.0ex\hbox{${\scriptscriptstyle \mathbf{m}}$}\mspace{-5.6mu}\mathbf{j}}}
\def\bjem{\mathbf{j}}
\def\baab{\mathbf{b}}
\def\bbaab{{\mathsurround 0pt\mbox{${\bf b}$\hspace*{-1.1ex}${\bf b}$}}}
\def\p{\partial}
\def\bp{\boldsymbol{\p}}
\def\unitc{\mathbf{1}}
\def\him{\imath}
\def\bhim{\boldsymbol{\imath}}
\def\invop#1{{#1^{\!\!-\!1}}}
\def\bHCLor{\boldsymbol{\Lambda}}

\def\bHCLorS{\comb{\boldsymbol{\circ}}{0.15ex}{0.24ex}{\boldsymbol{\Lambda}}}

\def\commut#1#2{\pmb{\bigl[}\, #1 \,\pmb{|}\, #2\, \pmb{\bigr]}}
\def\hconj#1{\mathchoice{{{}^{\boldsymbol{*}}\mspace{-2mu}#1}}{{{}^{\boldsymbol{*}}\mspace{-2mu}#1}}{{{}^{\boldsymbol{*}}\mspace{-4mu}#1}}{}{}}
\def\bRe{\boldsymbol{\Re}}
\def\bIm{\boldsymbol{\Im}}
\def\bDB{\mathbf{Y}}
\def\bEH{\mathbf{Z}}
\def\bnpc{\bnp^{\circ}}
\def\bnp{\comb{\boldsymbol{\cdot}}{0ex}{0.4ex} {\boldsymbol{\partial}}}
\def\nosum{\mathchoice{\textstyle \comb{\pmb{\bigl/}}{0ex}{1ex} {\sum}}{\comb{\pmb{\bigl/}}{0ex}{0.8ex} {\sum}}{}{}}
\def\prodf#1#2{\mathchoice{\pmb{\Bigl\langle} #1\, \pmb{|}\, #2 \pmb{\Bigr\rangle}}{\pmb{\bigl\langle} #1\, \pmb{|}\, #2 \pmb{\bigr\rangle}}{\pmb{\langle} #1\, \pmb{|}\, #2 \pmb{\rangle}}{}{}}
\def\bShST{\mathchoice{\combs{\rightarrow}{0.2ex}{0.5ex} \boldsymbol{\mathcal{J}}}{\combs{\rightarrow}{0.2ex}{0.5ex} \boldsymbol{\mathcal{J}}}{\combss{\rightarrow}{0.12ex}{0.2ex} \boldsymbol{\mathcal{J}}}{}{}}
\def\bRotS{\mathchoice{\comb{\circ}{0.05ex}{0.78ex} \boldsymbol{\mathcal{J}}}{\comb{\circ}{0.05ex}{0.78ex} \boldsymbol{\mathcal{J}}}{\combs{\circ}{0.05ex}{0.55ex} \boldsymbol{\mathcal{J}}}{}{}}
\def\Fourier#1{{}_{\mathtt{f}}\!\mspace{-1.5mu}#1}
\def\conjop#1{{}^{{\boldsymbol{\diamond}}}\!#1}
\def\cylaf#1#2{\mathchoice{\combs{\|}{0.42ex}{0.63ex}{\mathbf{C}^{#2}_{#1}}}{\combs{\|}{0.42ex}{0.63ex}{\mathbf{C}^{#2}_{#1}}}{\combss{\|}{0.245ex}{0.41ex}{\mathbf{C}^{#2}_{#1}}}{}{}}
\def\wavevp{\mathchoice{{\mathsurround 0pt\mbox{${k}$\hspace*{-1.05ex}${k}$}}}{{\mathsurround 0pt\mbox{${k}$\hspace*{-1.05ex}${k}$}}}{{\mathsurround 0pt\mbox{${\scriptstyle {k}}$\hspace*{-1.05ex}${\scriptstyle {k}}$}}}{{\mathsurround 0pt\mbox{${\scriptstyle {k}}$\hspace*{-1.15ex}${\scriptstyle {k}}$}}}{}}
\def\bwavevp{\mathchoice{{\mathsurround 0pt\mbox{$\mathbf{k}$\hspace*{-1.1ex}$\mathbf{k}$}}}{{\mathsurround 0pt\mbox{$\mathbf{k}$\hspace*{-1.1ex}$\mathbf{k}$}}}{{\mathsurround 0pt\mbox{${\scriptstyle \mathbf{k}}$\hspace*{-1.2ex}${\scriptstyle\mathbf{k}}$}}}{{\mathsurround 0pt\mbox{${\scriptstyle \mathbf{k}}$\hspace*{-1.3ex}${\scriptstyle\mathbf{k}}$}}}{}}
\def\Div{\mathnormal{\mathrm{Div}}}
\def\Curl{\mathnormal{\mathrm{Curl}}}
\def\norma#1{\bigl\|#1\bigr\|}
\def\sphersecf#1#2{\mathchoice{\comb{\circ}{0.24ex}{0.13ex}{\mathbf{S}_{#1}^{#2}}}{\comb{\circ}{0.24ex}{0.13ex}{\mathbf{S}_{#1}^{#2}}}{\combs{\circ}{0.16ex}{0.12ex}{\mathbf{S}_{#1}^{#2}}}{}{}}
\def\zc{\mathchoice{\combs{\circ}{0.15ex}{0.11ex}\mathrm{z}}{\combs{\circ}{0.15ex}{0.11ex}\mathrm{z}}{\combss{\circ}{0.05ex}{0.01ex}\mathrm{z}}{}{}}
\def\spherzonf#1#2#3{\mathchoice{\comb{\circ}{0.24ex}{0.23ex}{\mathbf{Z}_{#1#3}^{#2}}}{\comb{\circ}{0.24ex}{0.24ex}{\mathbf{Z}_{#1#3}^{#2}}}{\combs{\circ}{0.15ex}{0.2ex}{\mathbf{Z}_{#1#3}^{#2}}}{}{}}
\def\spheraf#1#2#3{\mathchoice{\comb{\circ}{0.23ex}{0.5ex}{\mathbf{C}^{#2#3}_{#1}}}{\comb{\circ}{0.23ex}{0.5ex}{\mathbf{C}^{#2#3}_{#1}}}{\combs{\circ}{0.17ex}{0.38ex}{\mathbf{C}^{#2#3}_{#1}}}{}{}}
\def\ClGord#1#2#3#4#5{\mathchoice{\combs{\otimes}{0.42ex}{0.25ex}{\mathrm{C}_{#4#5}^{#1#2#3}}}{\combs{\otimes}{0.42ex}{0.25ex}{\mathrm{C}_{#4#5}^{#1#2#3}}}{\combss{\otimes}{0.23ex}{0.1ex}{\mathrm{C}_{#4#5}^{#1#2#3}}}{}{}}
\def\RadfunC#1#2{\mathchoice{\combs{\boldsymbol{\uparrow}}{0.4ex}{0.55ex}{\mathbf{C}^{#1}_{#2}}}{\combs{\boldsymbol{\uparrow}}{0.4ex}{0.55ex}{\mathbf{C}^{#1}_{#2}}}{\combss{\boldsymbol{\uparrow}}{0.2ex}{0.35ex}{\mathbf{C}^{#1}_{#2}}}{}{}}
\def\RadfunS#1#2{\mathchoice{\combs{\boldsymbol{\uparrow}}{0.4ex}{0.2ex}{\mathbf{S}^{#1}_{#2}}}{\combs{\boldsymbol{\uparrow}}{0.4ex}{0.2ex}{\mathbf{S}^{#1}_{#2}}}{\combss{\boldsymbol{\uparrow}}{0.2ex}{0.1ex}{\mathbf{S}^{#1}_{#2}}}{}{}}
\def\garmcb#1#2#3#4{\mathchoice{{}_#3\!\combs{\combs{\|}{0ex}{0.27ex}{\approx}}{0.4ex}{0.41ex}{\mathbf{C}^{#4}_{#1#2}}}{{}_#3\!\combs{\combs{\|}{0ex}{0.27ex}{\approx}}{0.4ex}{0.41ex}{\mathbf{C}^{#4}_{#1#2}}}{{}_#3\!\combss{\combss{\|}{0.04ex}{0.17ex}{\approx}}{0.2ex}{0.25ex}{\mathbf{C}^{#4}_{#1#2}}}{}{}}
\def\garmsb#1#2#3#4{\mathchoice{{}_#2\!\combs{\comb{\circ}{-0.2ex}{0.06ex}{\approx}}{0.4ex}{0.41ex}{\mathbf{C}^{#3#4}_{#1}}}{{}_#2\!\combs{\comb{\circ}{-0.2ex}{0.06ex}{\approx}}{0.4ex}{0.41ex}{\mathbf{C}^{#3#4}_{#1}}}{{}_#2\!\combss{\combs{\circ}{-0.05ex}{0.13ex}{\approx}}{0.2ex}{0.25ex}{\mathbf{C}^{#3#4}_{#1}}}{}{}}

\begin{document}

\def\mtitle{BASIC SYSTEMS\\OF ORTHOGONAL FUNCTIONS\\ FOR SPACE-TIME MULTIVECTORS}

\def\mabstract{Space-time multivectors in Clifford algebra (space-time algebra) and their application to nonlinear electrodynamics are considered.
Functional product and infinitesimal operators for translation and rotation groups are introduced,
where unit pseudoscalar or hyperimaginary unit is used instead of imaginary unit.
Basic systems of orthogonal functions (plane waves, cylindrical, and spherical) for space-time multivectors are built by using
the introduced infinitesimal operators.
Appropriate orthogonal decompositions for electromagnetic field are presented.
These decompositions are applied to nonlinear electrodynamics.
Appropriate first order equation systems for cylindrical and spherical radial functions are obtained.
Plane waves, cylindrical, and spherical solutions to the linear electrodynamics
are represented by using the introduced orthogonal functions.
A decomposition of a plane wave in terms of the introduced spherical harmonics is obtained.}

\title{\textbf{\mtitle}}
\author{{\textbf{Alexander A. Chernitskii}}\\
\small Saint Petersburg Electrotechnical University\\
\small  Prof. Popov str. 5, St. Petersburg, 197376, Russia\\
\small  aa@cher.etu.spb.ru}
\date{}
\maketitle

{\small
\baselineskip=0.1ex
\tableofcontents
}
\newpage
\begin{abstract}
\mabstract
\end{abstract}

\section{Introduction}
\label{introd}
Multivectors are scalars, vectors, and fully asymmetric tensors.
The maximal rank of non-zero multivectors equals the dimension of space\footnote%
{If the number of tensor indices (the rank of tensor)
exceeds the dimension of space, then each tensor component has at least
two equal indices. Such fully asymmetric tensor is null tensor.}.
So called essential components of a fully asymmetric tensor define
all its components. In particular, such tensor of maximal rank has
one essential component: each its component is
{\em $\pm$some number} or $0$.

 Space-time multivectors are scalars, vectors, fully asymmetric
second-rank tensors or bivectors, fully asymmetric third-rank tensors or three-vectors or pseudovectors, and fully asymmetric fourth-rank
tensors or pseudoscalars. Bivector has six essential components and pseudovector has four ones.
There are $1+4+6+4+1=2^4=16$ essential components of all four-dimensional space-time multivectors.

Space-time multivectors have wide application for physics, in particular, for electrodynamics, where the electromagnetic field
is described as bivector space-time function.  Electromagnetic potential is space-time vector and dual potential
is pseudovector.

There is a very useful mathematical tool for manipulations with multivectors. This tool is based on Clifford algebra.
The dimension of Clifford algebra is equal to the quantity of all multivector essential components,
i.e. $16$ for four-dimensional space-time.
Having an appropriate multiplication table for non-commutative product
we can make algebraic manipulations with multivectors.
The members of the Clifford algebra are represented as hypercomplex numbers
and their non-commutative product is defined by the multiplication table.

The main object of this work is a building of basic systems of orthogonal
multivector functions and obtaining the multiplication tables for these functions.

\section{Space-time hypernumbers}
\label{hypnum}
Let us call hypercomplex numbers as hypernumbers.
The general form of space-time hypernumber is (see also my papers \cite{Chernitskii2002a,Chernitskii2003c})
\begin{equation}
\label{54511673}
\mathbf{C} =
\left(\unitc\,C^{0} {}+{} \bhim\,C^{\mathrm{IV}}\right)
{}+{}  \left(\unitc\,C^{\mathrm{I}}_\mu  {}+{}
\bhim\,C^{\mathrm{III}}_\mu\right)\baab^{\mu} {}+{}
\left(\unitc\,C^{\mathrm{II}\prime}_i {}+{} \bhim\,C^{\mathrm{II}\prime\prime}_i\right)\bbaab^i
\;\;,
\end{equation}
where $\unitc$ and $\bhim$ are hyperunit and hyperimaginary unit, $\baab^{\mu}$ are basis vectors
(Greek indices take on a value $0,1,2,3$),
$\bbaab^i$ are basis bivectors (Latin indices take on a value $1,2,3$),
and $C^{\mathrm{I}...\mathrm{IV}}_{...}$ are connected
with components of multivectors.

If we take $C^{\mathrm{I}}_\mu \equiv C^{\mathrm{III}}_\mu \equiv  0$
in (\ref{54511673}), then we have by definition an even space-time hypernumber.
On the contrary, if
$C^{0} \equiv C^{\mathrm{IV}} \equiv C^{\mathrm{II}\prime}_i \equiv
C^{\mathrm{II}\prime\prime}_i \equiv 0$, then the hypernumber is odd.

The first bracketed expression in (\ref{54511673}) will be called hyperscalar.

The expressions of type $\left(\unitc\,C^{\mathrm{I}}_\mu  {}+{} \bhim\,C^{\mathrm{III}}_\mu\right)$
will be called quasi-hyperscalars\footnote%
{The prefix ``quasi-'' can be omitted}. They differ from hyperscalars because of
existing tensor indices. Quasi-hy\-per\-sca\-lars are transform coupled with
coordinate system transformation: $\mathbf{C} = C_\mu\,\baab^{\mu} =
C^{\prime}_\mu\,\baab^{\prime\mu}$.
But a transformation of space-time rotation type for geometrical objects
leaves quasi-hyperscalars to be invariable: $\mathbf{C}^{\prime} = \bHCLor\,\mathbf{C}\,\invop{\bHCLor}
= \bHCLor\,C_\mu\,\baab^{\mu}\,\invop{\bHCLor} = C_\mu\,\bHCLor\,\baab^{\mu}\,\invop{\bHCLor}$,
where $\bHCLor$ is an even hypernumber realizing the space-time rotation (see, for example, \cite{HestenesSobczyk1984})
and $C_\mu$ are quasi-hyperscalars which are permutable with even hypernumbers (see late (\ref{48343708})).

Let us designate a hyperconjugate hypernumber as
\begin{equation}
\label{55221747}
\hconj{\mathbf{C}} \doteqdot \left(\unitc\,C^{0} {}-{} \bhim\,C^{\mathrm{IV}}\right)
{}+{}  \left(\unitc\,C^{\mathrm{I}}_\mu  {}-{}
\bhim\,C^{\mathrm{III}}_\mu\right)\baab^{\mu} {}+{}
\left(\unitc\,C^{\mathrm{II}\prime}_i {}-{} \bhim\,C^{\mathrm{II}\prime\prime}_i\right)\bbaab^i
\;\;.
\end{equation}
Hyperreal and hyperimaginary parts of hypernumber $\mathbf{C}$ (\ref{54511673}) are defined as
\begin{align}
\nonumber
\bRe\mathbf{C} &\doteqdot
\frac{1}{2}\left(\mathbf{C} + \hconj{\mathbf{C}}\right)
 = C^{0}\,\unitc {}+{}  C^{\mathrm{I}}_\mu  \,\baab^{\mu} {}+{} C^{\mathrm{II}\prime}_i\,\bbaab^i
\;\;,
\\
\bIm\mathbf{C} &\doteqdot
\frac{\bhim}{2}\left(\hconj{\mathbf{C} - \mathbf{C}}\right)
= C^{\mathrm{IV}}\,\unitc {}+{}  C^{\mathrm{III}}_\mu\,\baab^{\mu} {}+{} C^{\mathrm{II}\prime\prime}_i\,\bbaab^i
\;\;.
\label{40209010}
\end{align}

Hyperimaginary unit is the coordinate-free form of unit fully asymmetric fourth-rank tensor or space-time pseudoscalar%
\footnote{\label{footndes} Here I use designations some differing from used in my preceding papers \cite{Chernitskii2002a,Chernitskii2003c}.
All expressions of these my papers can be rewritten in the current designations by substitutions
$\varepsilon_{\mu\nu\rho\sigma}\to -\him_{\mu\nu\rho\sigma}$,
$\varepsilon^{\mu\nu\rho\sigma}\to -\him^{\mu\nu\rho\sigma}$, $\bhim\to -\bhim$.
}:
\begin{equation}
\label{56226711}
\bhim \doteqdot  \dfrac{1}{4!}\,
\him_{\mu\nu\rho\sigma}\baab^\mu\,\baab^\nu\,\baab^\rho\,\baab^\sigma
\;\;,
\end{equation}
where $\him_{\mu\nu\rho\sigma}$ are its components:
$\him_{0123} = -\sqrt{|\metr|}$, $\him^{0123} = \sqrt{|\metr|}^{-1}$, $\metr \doteqdot \mathrm{det} (\metr_{\mu\nu})$,
and $\metr_{\mu\nu}$ are components of metric tensor.

It is convenient to divide the non-commutative but associative and distributive hypernumber product to
symmetrical and asymmetrical parts:
\begin{subequations}
\label{69905217}
\begin{align}
\label{44218146}
&\st{1}{\mathbf{C}} \, \st{2}{\mathbf{C}} = \st{1}{\mathbf{C}} \bcdot \st{2}{\mathbf{C}} + \st{1}{\mathbf{C}} \bwedge \st{2}{\mathbf{C}}
\;\;,\\
\label{64298295}
&\st{1}{\mathbf{C}} \bcdot \st{2}{\mathbf{C}} \doteqdot
\dfrac{1}{2}\Bigl(\st{1}{\mathbf{C}}\,\st{2}{\mathbf{C}} + \st{2}{\mathbf{C}}\,\st{1}{\mathbf{C}}\Bigr)
\;\;,\quad
\st{1}{\mathbf{C}} \bwedge \st{2}{\mathbf{C}} \doteqdot
\dfrac{1}{2}\Bigl(\st{1}{\mathbf{C}}\,\st{2}{\mathbf{C}} - \st{2}{\mathbf{C}}\,\st{1}{\mathbf{C}}\Bigr)
\;\;.
\end{align}
\end{subequations}
Introduced operations $\bcdot$ and $\bwedge$ will be called symmetrical and asymmetrical products\footnote%
{The designations used here  for symmetrical and asymmetrical products
($\bcdot$ and $\bwedge$) differs from ones used in my preceding papers
\cite{Chernitskii2002a,Chernitskii2003c} ($\cdot$ and $\wedge$).
Customary meanings of the symbols $\cdot$ and $\wedge$ are internal and exterior products.
Really we must distinguish the symmetrical and asymmetrical products
from internal and exterior ones.
There is the coincidence of these product pairs for vectors
(it is clear, also for scalars) but not for the general case.
The use of symmetrical and asymmetrical products for hypernumbers is
preferably, because the definition of these products by using
non-commutative one (\ref{64298295}) is independent of types of
multiplied multivectors, in contrast to appropriate
definitions for internal and exterior products.
All expressions of my papers \cite{Chernitskii2002a,Chernitskii2003c} can be rewritten in the current designations by substitutions
$\cdot\to\bcdot$ and $\wedge\to\bwedge$.
}.
They are non-associative but distributive.
Consecutive use of these operations without brackets, such that
$\st{1}{\mathbf{C}}\bcdot\st{2}{\mathbf{C}}\bcdot\st{3}{\mathbf{C}}$
or $\st{1}{\mathbf{C}}\bwedge\st{2}{\mathbf{C}}\bwedge\st{3}{\mathbf{C}}$, must be considered as symmetrization or
alternation respectively by all co-factors. For example
\begin{align}
\nonumber
\st{1}{\mathbf{C}} {}\bwedge{} \st{2}{\mathbf{C}} {}\bwedge{} \st{3}{\mathbf{C}}
&\doteqdot \frac{1}{3!}\Bigl(\st{1}{\mathbf{C}}\,\st{2}{\mathbf{C}}\,\st{3}{\mathbf{C}} -\st{1}{\mathbf{C}}\,\st{3}{\mathbf{C}}\,\st{2}{\mathbf{C}} + \st{3}{\mathbf{C}}\,\st{1}{\mathbf{C}}\,\st{2}{\mathbf{C}}
- \st{2}{\mathbf{C}}\,\st{1}{\mathbf{C}}\,\st{3}{\mathbf{C}} + \st{2}{\mathbf{C}}\,\st{3}{\mathbf{C}}\,\st{1}{\mathbf{C}} - \st{3}{\mathbf{C}}\,\st{2}{\mathbf{C}}\,\st{1}{\mathbf{C}} \Bigr)
\\
&\neq \st{1}{\mathbf{C}} {}\bwedge{} \Bigl(\st{2}{\mathbf{C}} {}\bwedge{} \st{3}{\mathbf{C}}\Bigr) \neq
\Bigl(\st{1}{\mathbf{C}} {}\bwedge{} \st{2}{\mathbf{C}}\Bigr) {}\bwedge{} \st{3}{\mathbf{C}}
\;\;.
\label{ppoiejdd}
\end{align}

Using designations (\ref{64298295}), let us write the following multiplication table
(see also \cite{Chernitskii2002a,Chernitskii2003c}):
\begin{align}
\nonumber
&\mathbf{C}\,\unitc = \unitc\,\mathbf{C} = \mathbf{C}
\;,\quad
\bhim\,\bhim = -\unitc
\;,\quad
\bhim\,\baab^{\mu} = -\baab^{\mu}\,\bhim\;,\quad \bhim\,\bbaab^i = \bbaab^i\,\bhim\;,
\\
\nonumber
&\baab^\mu \bcdot \baab^\nu = \unitc \,\metr^{\mu\nu}\;,\quad
\baab^i {}\bwedge{} \baab^0 = \bbaab^i\;,\quad
\baab^i {}\bwedge{} \baab^j = \him^{0ijl}\,\bhim\,\bbaab_l\;,\quad \baab_\nu = \metr_{\nu\mu}\,\baab^\mu\;,
\\
\nonumber
&\baab^0 {}\bcdot{} \bbaab^i  = 0\;,\quad \baab^i {}\bcdot{} \bbaab^j = -\him^{0ijk}\,\bhim\,\baab_k\;,
\quad
\baab^i {}\bwedge{} \bbaab_j = -\delta^i_j\,\baab_0\;,\quad
\baab^0 {}\bwedge{} \bbaab_i = \baab_i\;,
\\
\nonumber
&\bbaab^i \bcdot \bbaab^j = -\metr^{00}\,\metr^{ij} {}+{} \metr^{i0}\,\metr^{j0}\;,\quad
\bbaab^i {}\bwedge{} \bbaab_j = \baab^i {}\bwedge{} \baab_j\;,
\\
&\bbaab^i = \left(-\metr^{00}\,\metr^{ij} {}+{} \metr^{i0}\,\metr^{j0}\right)\bbaab_j
\;.
\label{48343708}
\end{align}

There are so called zero divisors in this hypercomplex system.
Zero divisor is a hypernumber $\mathbf{C}$ such that $\mathbf{C}\,\mathbf{X} =0$ (left) or $\mathbf{X}\,\mathbf{C} =0$
(right) for some hypernumber $\mathbf{X} \neq 0$.
There is no inverse element for zero divisor. Really, if $\mathbf{C}\,\mathbf{X} = 0$, then
$\invop{\mathbf{C}}\,\mathbf{C} = \unitc\,\Rightarrow\,\invop{\mathbf{C}}\,\mathbf{C}\,\mathbf{X} = \mathbf{X}%
\,\Rightarrow\,0 = \mathbf{X}$. For example, if $(\bbaab_1)^2 = \unitc$, then
$(\unitc - \bbaab_1)\,(\unitc + \bbaab_1) = \unitc - (\bbaab_1)^2 = 0$.

It should be noted that the coefficients $C^{\mathrm{I}...\mathrm{IV}}_{...}$ suppose to be real numbers.
Thus there is not the imaginary unit in this hypernumber system.
But the hyperimaginary unit $\bhim$ can be used as customary imaginary unit for
even space-time hypernumbers, because of $\bhim\,\bhim = -\unitc$ and $\bhim\,\bbaab^i = \bbaab^i\,\bhim$.

The symmetrical and asymmetrical products for bivectors corresponds
with scalar and vector products respectively for space vectors with
quasi-hyperscalar components:
\begin{subequations}
\label{41729628}
\begin{align}
\label{41781392}
&\st{1}{\mathbf{C}}\bcdot\st{2}{\mathbf{C}} =
\st{1}{\mathbf{C}}\cdot\st{2}{\mathbf{C}} =
\metr^{ij}\,\st{1}{{C}}_i\,\st{2}{{C}}_j\,\unitc
\;\;,\quad
\\
&\st{1}{\mathbf{C}}\bwedge\st{2}{\mathbf{C}} = \bhim\,\st{1}{\mathbf{C}}\times\st{2}{\mathbf{C}}
\;\;,\quad
\st{1}{\mathbf{C}}\times\st{2}{\mathbf{C}} \doteqdot
\bigl(\him^{0ijk}\,\st{1}{{C}}_j\,\st{2}{{C}}_k\bigr)\,\bbaab_{i}
\;\;,
\label{72877710}
\end{align}
\end{subequations}
where $\st{q}{\mathbf{C}} = \st{q}{{C}}_i\,\bbaab^{i}$, $\st{q}{C}_i$ are quasi-hyperscalars, and
it is supposed $\metr_{00}=-1$, $\metr_{i0}=0$.

Since customary complex numbers do not use in the mathematical tool
under consideration, we can, in principle, omit the prefix ``hyper''
in some words.
But at the present paper I use the original long words.

\section{Functional product}
\label{funcprod}
Because the hyperimaginary unit $\bhim$ is permutable with even space-time hypernumbers, any manipulations with these
hypernumbers are more convenient than for odd ones. But odd hypernumber can be transformed to even one with multiplication by
the basis vector $\baab_0$. Thus we will build basic systems of orthogonal functions for even hypernumbers.

Let us define functional product for hyperscalar or bivector space-time functions:
\begin{equation}
\label{kkklikdjjff}
\prodf{\st{1}{\mathbf{C}}}{\st{2}{\mathbf{C}}} \equiv
\int\limits_{\Vol}
\bigl(\st{1}{\mathbf{C}}\bcdot\hconj{\st{2}{\mathbf{C}}}\bigr)
\;\dVol
\;\;,
\end{equation}
where $\Vol$ is some space-time volume and $\dVol$ is its element,
$\st{1}{\mathbf{C}}(\mathbf{x})$ and $\st{2}{\mathbf{C}}(\mathbf{x})$ are both hyperscalar or bivector functions.
The functional product for a hyperscalar function with a bivector one is zero by definition.
The functional product takes constant hyperscalar or quasi-hyperscalar values.

It is evident that the functional product defined as (\ref{kkklikdjjff}) has all traditional properties of
functional product defined for conventional complex functions.

Let us define the following norm of functional vector:
\begin{equation}
\label{41825843}
\norma{\st{1}{\mathbf{C}}} \doteqdot \sqrt{\prodf{\st{1}{\mathbf{C}}}{\st{1}{\mathbf{C}}}}
\;\;.
\end{equation}

Also we us define the module of hyperscalar or bivector:
\begin{equation}
\label{34684530}
\left|\mathbf{C}\right| \doteqdot
\sqrt{\mathbf{C}\bcdot\hconj{\mathbf{C}}}
\;\;.
\end{equation}

Adjoint operator is defined in the regular way:
\begin{equation}
\label{ddjekks}
\prodf{\mathbf{Q}\,\st{1}{\mathbf{C}}}{\st{2}{\mathbf{C}}} \doteqdot
\prodf{\st{1}{\mathbf{C}}}{\conjop{\mathbf{Q}}\,\st{2}{\mathbf{C}}}
\;\;.
\end{equation}

Eigenvalues for operators take on quasi-hyperscalar values.

For self-adjoint ($\conjop{\mathbf{Q}}=\mathbf{Q}$) and anti-self-adjoint ($\conjop{\mathbf{Q}}=-\mathbf{Q}$) operators
there are two useful properties:
its eigenvalues are hyperreal and hyperimaginary accordingly
and its eigenfunctions with different eigenvalues are orthogonal.
The appropriate proof is very simple, it is fully analogous to the
case of functional product defined for conventional complex
functions.

\section[Infinitesimal operators]{Infinitesimal operators\\ for translation and rotation groups}
\label{infop}
A self-adjoint infinitesimal shift operators has the form
\begin{equation}
\label{sskkeiir}
 \bShST_\mu\doteqdot -\bhim\,\frac{\p}{\p x^\mu}
\;\;.
\end{equation}
We have the appropriate invariant self-adjoint infinitesimal operator:
\begin{equation}
\label{eewlllmkjf}
\bShST_\mu\, \bShST^\mu = -\frac{\p^2}{\p x^\mu\, \p x_\mu}
\;\;.
\end{equation}

A self-adjoint infinitesimal rotation operators can be obtained in space-time algebra formalism.
These operators have the form\footnote%
{See also the accordance between $\bwedge$ and $\times$ operations
for the case of bivector functions (\ref{72877710}).}
\begin{equation}
\label{kkjnmfndd}
\bRotS^i \doteqdot -\bhim\,\him^{0ijk}\,x_j\,\frac{\p}{\p x^k} + \bbaab^i\bwedge
\;\;.
\end{equation}

Thus we have also the appropriate anti-self-adjoint operators $-\bhim\,\bShST_\mu$ and $-\bhim\,\bRotS^i$.
These operators realize group infinitesimal transformations. For example, we have
the infinitesimal shift and rotation
for some space-time hypernumber function $\mathbf{C} (x^0,x^1,x^2,x^3)$ about the
axis $x^3$:
\begin{subequations}
\label{aakmnff}
\begin{align}
\nonumber
\mathbf{C}^{\prime} &= \mathbf{C} (x^0,x^1,x^2,x^3 - \delta a) = \mathbf{C} (x^0,x^1,x^2,x^3) - \delta a\,\frac{\p \mathbf{C}}{\p x^3}
\\
\label{aallkmmdd}
&= \left(\unitc -\delta a\,\bhim\,\bShST_3\right)\mathbf{C}
\;\;,\\
\nonumber
\mathbf{C}^{\prime\prime} &=
\mathbf{C} (x^0,x^1 + x_2\,\delta\varphi,x^2- x_1\,\delta\varphi,x^3) + \mathbf{C}\bwedge\left(\bhim\,\bbaab^3\right)\delta\varphi
\\
\label{ooolder}
&= \left(\unitc -\delta\varphi\,\bhim\,\bRotS^3\right)\mathbf{C}
\;\;.
\end{align}\end{subequations}

If it is possible to permute the order of differentiation, that is outside of singular sets, then we can obtain the following
customary commutation relations for the self-adjoint infinitesimal operators:
\begin{equation}
\label{39747725}
\commut{\bShST_\mu}{\bShST_\nu} = 0
\;\;,\qquad
\commut{\bRotS^i}{\bRotS^j} =\him^{0ijk}\,\bhim\,\bRotS_k
\;\;.
\end{equation}

In the regular way (see, for example,
\cite{GelfMilnShap1958e,ElliottDawber1979v1})
we obtain the appropriate raising and reducing operators
\begin{equation}
\label{wwsderggv}
\bRotS^{+} = \bRotS^1 + \bhim\,\bRotS^2
\quad,\qquad
\bRotS^{-} = \bRotS^1 - \bhim\,\bRotS^2
\;\;.
\end{equation}
with the customary commutation relations
\begin{equation}
\label{40959768}
\commut{\bRotS^{+}}{\bRotS^{3}} = - \bRotS^{+}
\;\;,\quad
\commut{\bRotS^{-}}{\bRotS^{3}} =  \bRotS^{-}
\;\;,\quad
\commut{\bRotS^{+}}{\bRotS^{-}} =  2\,\bRotS^{3}
\;\;.
\end{equation}

We have the invariant self-adjoint infinitesimal operator for rotation
\begin{subequations}
\label{40684653}
\begin{align}
\label{41419070}
(\bRotS)^2 &\doteqdot (\bRotS_1)^2 + (\bRotS_2)^2 + (\bRotS_3)^2
\\
\label{40047342}
 &= \bRotS_{-}\,\bRotS_{+} + (\bRotS_3)^2 + \bRotS_3 = \bRotS_{+}\,\bRotS_{-} + (\bRotS_3)^2 - \bRotS_3
\end{align}
\end{subequations}
and the appropriate commutation relations
\begin{equation}
\label{41241258}
\commut{(\bRotS)^2}{\bRotS_{j}} = \commut{(\bRotS)^2}{\bRotS_{-}} = \commut{(\bRotS)^2}{\bRotS_{+}} = 0
\;\;.
\end{equation}

\section{Electrodynamics}
\label{electrod}
Two antisymmetric tensors or bivectors of electromagnetic field are represented
in the form (see my paper \cite{Chernitskii2002a})
\begin{subequations}
\label{HyperEMF}
\begin{align}
\bFem &\doteqdot \frac{1}{2}\,\Fem_{\mu\nu}\,\baab^{\mu}\,\baab^{\nu} {}={}
\Eem_i\,\bbaab^{i} {}+{} \Bem^i\,\bhim\bbaab_{i}
= \bEem  {}+{} \bhim\,\bBem
\;\;,
\label{Def:bFem}
\\
\bfem &\doteqdot \frac{1}{2}\,\fem^{\mu\nu}\,\baab_{\mu}\,\baab_{\nu} {}={}
\Dem^i\,\bbaab_{i} {}+{} \Hem_i\,\bhim\bbaab^{i}
= \bDem  {}+{} \bhim\,\bHem
\;\;,
\label{Def:bfem}
\end{align}
\end{subequations}
where $\bEem$ and $\bHem$ are electric and magnetic field intensities,
$\bDem$ and $\bBem$ are electric and magnetic inductions.

A hypernumber form of nonlinear electrodynamics was obtained in my work \cite{Chernitskii2002a}. Let us write
this equation with electromagnetic current (see \cite{Chernitskii2003c}):
\begin{equation} \label{79601661}
\bp \bcdot\bFem  {}+{} \bp \bwedge\bfem = -4\pi\,\bjem
\quad,
\end{equation}
where
\begin{equation}
\label{defbp}
\bp \doteqdot  \baab^{\mu}\,\p_\mu {\,}\doteqdot{\,}
\baab^{\mu}\,\dfrac{\p}{\p x^\mu}
\quad.
\end{equation}
Here the operator of coordinate differentiation $\p_\mu$ is considered as scalar, it operate only to expression being on the right.

It is convenient to introduce the following two quasi-bivectors of elec\-tro\-mag\-ne\-tic induction and intensity
(see also my paper \cite{Chernitskii1999}):
\begin{equation}
\label{YZ}
\bDB \doteqdot \bDem + \bhim\,\bBem
\;\;,\quad
\bEH \doteqdot \bEem + \bhim\,\bHem
\;\;.
\end{equation}
The quasi-bivectors are invariant only for space transformations of coordinate system but not for transformations
affecting time.

Let us multiply equation (\ref{79601661}) by $\baab_0$ on the left.
Using designations (\ref{YZ}) and taking into consideration (\ref{HyperEMF}) and
(\ref{48343708}), we obtain the following equation for elec\-tro\-mag\-ne\-tic quasi-bivectors:
\begin{subequations}
\label{44396100}
\begin{align}
\label{dkmfjysg}
& \p_0\,\bDB + \bnp \bcdot \bDB + \bnp\bwedge\bEH = -4\pi\,\baab_0\,\bjem
\;\;,
\\
\label{44429658}
& \p_0\,\bDB + \Div\bDB + \bhim\,\Curl\bEH = -4\pi\,\baab_0\,\bjem
\;\;,
\end{align}
\end{subequations}
where $\p_\mu\doteqdot\dfrac{\p}{\p x^\mu}$ and
\begin{equation}
\label{wkdhjfof}
\bnp\doteqdot\baab_0\,\baab^i\,\p_i
\end{equation}
($i=1,2,3$)  is an operator of space differentiation and
\begin{equation}
\label{42954357}
\Div\bDB \doteqdot \bnp \bcdot \bDB
\quad,\qquad
\Curl\bEH \doteqdot -\bhim\,\bnp \bwedge \bEH \doteqdot \bnp\times \bEH
\quad.
\end{equation}
For coordinate systems with $\metr_{00}=-1$ and $\metr_{0i} = 0$ we have $\bnp= \bbaab^{i}\,\p_i$.

As we can see, the hyperimaginary unit $\bhim$ is permutable with $\p_0$, $\bDB$, $\bEH$, $\bnp$, and $\baab_0\,\bjem$,
 which are contained into equation (\ref{44396100}).
Thus the hyperimaginary unit can be used here as conventional imaginary unit.

It is evident that $\p_0$ is anti-self-adjoint operator about functional product (\ref{kkklikdjjff}).
Using multiplication table (\ref{48343708}) we can verify directly for flat space-time that
$\bnp$ is also anti-self-adjoint operator for even hypernumber functions, i.e. $\conjop{\bnp} = -\bnp$. Thus
$\pm \bhim\,\p_0$ and $\pm \bhim\,\bnp$ are self-adjoint operators.

In addition to equations (\ref{79601661}) or (\ref{44396100}) we must have relations connecting $\bFem$ with $\bfem$ or
$\bDB$ with $\bEH$ respectively. They can be called constitutive relations for the general case including
also nonlinear vacuum electrodynamics (see my articles \cite{Chernitskii1999,Chernitskii2004a})
and electrodynamics in medium.

In particular, for the case of linear vacuum or simplest constitutive relations
we have $\bFem=\bfem=\bDB=\bEH$,
and we can write instead of (\ref{dkmfjysg}) the following equation:
\begin{equation}
\label{48931593}
\p_0\,\bDB + \bnp \,\bDB  = -4\pi\,\baab_0\,\bjem
\;\;.
\end{equation}

\section{Plane waves}
\label{planewaves}
Eigenfunctions of invariant self-adjoint operator (\ref{eewlllmkjf}) are
$\exp {(\bhim\,k_\mu\,x^\mu)}$.
Representation for even hypernumber function in terms of this system of orthogonal functions have the form:
\begin{equation}
\label{ssewwdf}
\mathbf{Q} (\mathbf{x}) = \int\limits_{\Fourier{\Vol}} \Fourier{\mathbf{Q}}\,\e{\bhim\,k_\mu\,x^\mu}\,\df\Fourier{\Vol}
\;\;,
\end{equation}
where $\Fourier{\Vol}$ is unlimited volume in space of wave vectors, $\df\Fourier{\Vol}$  is its element,
 $\Fourier{\mathbf{Q}}$ is the Fourier transform for the function $\mathbf{Q} (\mathbf{x})$:
\begin{equation}
\label{wwqsder}
\Fourier{\mathbf{Q}}(\mathbf{k}) =
\frac{1}{(2\pi)^{4}}\,\prodf{{\mathbf{Q}}(\mathbf{x})}{\e{\bhim\,k_\mu\,x^\mu}}
= \frac{1}{(2\pi)^4}\int\limits_{\Vol} {\mathbf{Q}}(\mathbf{x})\,\e{-\bhim\,k_\mu\,x^\mu}\,\dVol
\;\;,
\end{equation}
where $\Vol$ is unlimited space-time volume.

Let us substitute representation of type (\ref{ssewwdf}) for $\bDB$ and $\bEH$ into equation (\ref{44396100})
without right-hand part or current. We have:
\begin{equation}
\label{46127686}
\bwavevp\bcdot\Fourier{\bDB} + \bwavevp\bwedge\Fourier{\bEH} = \omega\,\Fourier{\bDB}
\;\;,
\end{equation}
where $\omega \doteqdot -k_0$ is cyclic frequency, $\bwavevp \doteqdot \baab_0\,\baab^i\,k_i$ is space wave vector.

Hyperscalar part of equation (\ref{46127686}) gives the condition of transverse waves $\bwavevp\bcdot\Fourier{\bDB} = 0$.

For the case of simplest constitutive relations  $\bDB=\bEH$ we have
\begin{equation}
\label{42747689}
\bwavevp\,\Fourier{\bDB} = \omega\,\Fourier{\bDB}
\;\;.
\end{equation}
Since the hyperreal bivector $\bwavevp$ have inverse one $\invop{\bwavevp} = \bwavevp/\wavevp^2$,
where $\wavevp\doteqdot |\bwavevp|$, we also obtain from (\ref{42747689})
that $\Fourier{\bDB} = \omega\,\invop{\bwavevp}\, \Fourier{\bDB}$. Combining this relation with (\ref{42747689})
we have
\begin{equation}
\label{47176287}
\omega\,\invop{\bwavevp}\, \Fourier{\bDB} = \frac{\bwavevp}{\omega}\,\Fourier{\bDB}
\quad\Longrightarrow\quad
\omega\,\Fourier{\bDB} = \frac{\bwavevp\,\bwavevp}{\omega}\,\Fourier{\bDB}
\quad\Longrightarrow\quad
\omega^2 = \wavevp^2
\;\;.
\end{equation}

\section{Cylindrical waves}
\label{cylindricalwaves}
Let us consider a cylindrical coordinate system $\{x^0,\,\rho,\,\varphi,\,x^3\}$ in flat space-time
($x^0 = -x_0$, $x^i = x_i$):
$x_1 = \rho\,\cos\varphi$,
$x_2 = \rho\,\sin\varphi$.

\begin{Def}
\label{46403691}
The angle cylindrical functions are called the
even space-time hypernumber
eigenfunctions depending on polar angle $\varphi$
for the operator $\bRotS_{3}$ and are denoted by $\cylaf{j}{m}$:
\begin{equation}
\label{44146270}
\bRotS_3\,\cylaf{j}{m} = m\,\cylaf{j}{m}
\;\;.
\end{equation}
 where $\cylaf{j}{m}= \cylaf{j}{m}(\varphi)$.
\end{Def}

Infinitesimal operator for space rotation (\ref{kkjnmfndd}) about $x_3$ axis can be written in the form:
\begin{equation}
\label{ddlfkgjnneexs}
\bRotS_{3} = -\bhim\,\frac{\p}{\p\varphi} + \bbaab_{3}\bwedge
\;\;.
\end{equation}

There are the following hyperscalar and bivector angle cylindrical functions:
\begin{subequations}
\label{40806337}
\begin{align}
\label{aaseerfvgg}
&\cylaf{\mathrm{s}}{m} = \e{\bhim\,m\,\varphi}
\;\;,\\
&\label{12342156}
\cylaf{1}{m}=\e{\bhim\,(m-1)\,\varphi}\,\bbaab_{+}
\;\;,\quad
\cylaf{-1}{m}=\e{\bhim\,(m+1)\,\varphi}\,\bbaab_{-}
\;\;,\quad
\cylaf{0}{m}=\e{\bhim\,m\,\varphi}\,\bbaab_{3}
\;\;,
\end{align}
\end{subequations}
where we take
 $j=\mathrm{s}$ for hyperscalar and $j=0,\pm 1$ for bivectors,
\begin{equation}
\label{qqqwddtyh}
 \bbaab_{+}\doteqdot \bbaab_{1} + \bhim\,\bbaab_{2}
\quad,\qquad
\bbaab_{-}\doteqdot \bbaab_{1} - \bhim\,\bbaab_{2}
\quad.
\end{equation}
According to (\ref{48343708}) and (\ref{qqqwddtyh}) we have the following multiplication table for these bivectors
(metric tensor is pseudo-Euclidean):
\begin{subequations}
\label{wwplljjf}
\begin{align}
\label{ddkiuuygtg}
&  \bbaab_{+}\,\bbaab_{+} = \bbaab_{-}\,\bbaab_{-} = 0
\;\;,\quad
\bbaab_{+}\bcdot\bbaab_{-} = 2\cdot\unitc
\;\;,\quad
\bbaab_{+}\bwedge\bbaab_{-} = 2\,\bbaab_3
\;\;,\\
\nonumber
 &  \bbaab_{+}\bcdot\bbaab_1 = \bbaab_{-}\bcdot\bbaab_1 = \unitc
\quad,\qquad
 \bbaab_{+}\bcdot\bbaab_2 = -\bbaab_{-}\bcdot\bbaab_2 = \bhim
\;\;,\\
\label{ddkiuuywwwgtg}
& \bbaab_{+}\bcdot\bbaab_3 = \bbaab_{-}\bcdot\bbaab_3 = 0
\;\;,\\
\nonumber
& \bbaab_{+}\bwedge\bbaab_1 = -\bbaab_{-}\bwedge\bbaab_1 = \bbaab_3
\;\;,\quad
\bbaab_{+}\bwedge\bbaab_2 = \bbaab_{-}\bwedge\bbaab_2 = \bhim\,\bbaab_3
\;\;,\\
& \bbaab_{+}\bwedge\bbaab_3 = -\bbaab_{+} \quad,\qquad  \bbaab_{-}\bwedge\bbaab_3 = \bbaab_{-}
\;\;.
\end{align}
\end{subequations}
As we see in (\ref{ddkiuuygtg}), bivectors $\bbaab_{-}$ and $\bbaab_{+}$ have zero squares.
But since $\hconj{\bbaab_{-}} = \bbaab_{+}$ and $\hconj{\bbaab_{+}} = \bbaab_{-}$, six bivectors
$\bbaab_{-}$, $\bbaab_{+}$, $\bbaab_{3}$, $\bhim\,\bbaab_{-}$, $\bhim\,\bbaab_{+}$, $\bhim\,\bbaab_{3}$ are bivector basis
about functional product (\ref{kkklikdjjff}).

For representation an arbitrary even hypernumber space-time function let us consider also plane waves propagating along
$x^3$ axis. Thus we have:
\begin{equation}
\label{sqasdsewwdf}
\mathbf{Q} (x^0,\rho,\varphi,x^3) =
\sum_{j}^{\mathrm{s},0,1,-1}\sum_{m=-\infty}^{\infty}\cylaf{j}{m}\iint\limits_{-\infty}^{\infty}
\Fourier{\mathbf{Q}_j^m}\,\e{\bhim\,(k_3\,x^3 - \omega\,x^0)}\,\df \omega \df k_3
\;\;,
\end{equation}
where
$\Fourier{\mathbf{Q}_j^m}$ are quasi-hyperscalar functions:
\begin{equation}
\label{wwqsderder}
\Fourier{\mathbf{Q}_j^m}(\rho,\omega,k_3) =
\frac{1}{(2\pi)^{2}\,\norma{\cylaf{j}{m}}^2}\,
\prodf{\mathbf{Q} (x^0,\rho,\varphi,x^3)}{\cylaf{j}{m}\,\e{\bhim\,(k_3\,x^3 - \omega\,x^0)}}_{\varphi x^3 x^0}
\;\;.
\end{equation}
Here the limits on integrals in functional product are
$0< \varphi < 2\pi$,
$-\infty <x^3 < \infty$, and $-\infty < x^0 < \infty$,
there is not a summation by $j$ index, $\norma{\cylaf{1}{m}}^2 = \norma{\cylaf{-1}{m}}^2 = 4\pi$,  $\norma{\cylaf{0}{m}}^2 = 2\pi$.

The operator of space differentiation $\bnp$ (\ref{wkdhjfof}) has the following form in cylindrical coordinate system:
\begin{equation}
\label{wlllrokjds}
\bnp = \bbaab^{\rho}\,\frac{\p}{\p \rho} + \bbaab^{\varphi}\,\frac{\p}{\p \varphi} + \bbaab^{3}\,\frac{\p}{\p x^3}
\doteqdot \bnp^{\rho} + \bnp^{\varphi} + \bnp^{3}
\;\;,
\end{equation}
where $\bbaab^{\rho}\doteqdot \baab^{\rho}\bwedge\baab^0 = \bbaab_{\rho}$, $\bbaab^{\varphi}\doteqdot \baab^{\varphi}\bwedge\baab^0$.
In view of the metric for cylindrical coordinate system ($\metr_{\rho\rho} = \metr_{33}= 1$, $\metr_{\varphi\varphi}=\rho^2$) we have
$\bbaab_{\varphi} = \rho^2\,\bbaab^{\varphi}$.
Being guided by geometrical consideration, multiplication table (\ref{wwplljjf}),  and definition (\ref{12342156}) we have
\begin{subequations}
\label{38920527}
\begin{align}
\label{38246525}
 & \bbaab_{\rho} = \bbaab_{1}\,\cos\varphi +\bbaab_{2}\,\sin\varphi =
\frac{1}{2}\left(\cylaf{-1}{0} + \cylaf{1}{0}\right)
\;\;,
\\
\label{38253743}
 & \bbaab^{\varphi} = \frac{1}{\rho}\left(-\bbaab_{1}\,\sin\varphi +\bbaab_{2}\,\cos\varphi\right) =
\frac{\bhim}{2\,\rho}\left(\cylaf{-1}{0} - \cylaf{1}{0}\right)
\;\;,
\\
\label{40253979}
&\bbaab_{3} = \cylaf{0}{0}
\;\;.
\end{align}
\end{subequations}

According to (\ref{40806337}), (\ref{wwplljjf}) there is the following multiplication and hy\-per\-con\-ju\-ga\-tion table
for the cylindrical angle functions:
\begin{align}
\nonumber
&  \cylaf{1}{m}\,\cylaf{1}{n} = \cylaf{-1}{m}\,\cylaf{-1}{n} = 0
\;\;,\quad
\cylaf{1}{m}\bcdot\cylaf{0}{n} = \cylaf{-1}{m}\bcdot\cylaf{0}{n} = 0
\;\;,\\
\nonumber
& \cylaf{1}{m}\bcdot\cylaf{-1}{n} = 2\,\cylaf{\mathrm{s}}{m+n}
\;\;,\quad
\cylaf{0}{m}\bcdot\cylaf{0}{n} = \cylaf{\mathrm{s}}{m+n}
\;\;,\quad
\cylaf{j}{m}\,\cylaf{\mathrm{s}}{n} = \cylaf{\mathrm{s}}{n}\,\cylaf{j}{m} = \cylaf{j}{m+n}
\;\;,\\
\nonumber
& \cylaf{1}{m}\bwedge\cylaf{-1}{n} = 2\,\cylaf{0}{m+n}
\;\;,\quad
\cylaf{1}{m}\bwedge\cylaf{0}{n} = -\cylaf{1}{m+n} \;\;,\quad  \cylaf{-1}{m}\bwedge\cylaf{0}{n} = \cylaf{-1}{m+n}
\;\;,
\\
\label{40948743}
&
\hconj{\cylaf{\mathrm{s}}{m}} = \cylaf{\mathrm{s}}{-m}
\;\;,\quad
\hconj{\cylaf{0}{m}} = \cylaf{0}{-m}
\;\;,\quad
\hconj{\cylaf{1}{m}} = \cylaf{-1}{-m}
\;\;,\quad
\hconj{\cylaf{-1}{m}} = \cylaf{1}{-m}
\;\;.
\end{align}

Taking into consideration (\ref{38253743}) and (\ref{40948743}) we can obtain the following table for application of the
angle differentiation operator $\bnp_\varphi \doteqdot\bbaab^{\varphi}\,\p_\varphi$ to the cylindrical angle functions:
\begin{align}
\nonumber
2\,\rho\,\bnp^{\varphi}\bcdot \cylaf{\mathrm{s}}{m} &= m\,\bhim\left(\cylaf{-1}{m} - \cylaf{1}{m}\right)
\;\;,&\quad
\bnp^{\varphi}\bwedge \cylaf{\mathrm{s}}{m} &= 0
\;\;,
\\
\nonumber
\bnp^{\varphi}\bcdot \cylaf{0}{m} &= 0
\;\;,&\quad
2\,\rho\,\bnp^{\varphi}\bwedge \cylaf{0}{m} &= -m\,\left(\cylaf{-1}{m} + \cylaf{1}{m}\right)
\;\;,
\\
\rho\,\bnp^{\varphi}\bcdot \cylaf{\pm 1}{m} &= (1\mp m)\,\cylaf{\mathrm{s}}{m}
\;\;,&\quad
\rho\,\bnp^{\varphi}\bwedge \cylaf{\pm 1}{m} &= (m\mp 1)\,\cylaf{0}{m}
\label{40791520}
\;\;.
\end{align}

Let us substitute representation of type (\ref{sqasdsewwdf}) for two electromagnetic quasibivectors $\bDB$ and $\bEH$ (\ref{YZ})
into equation (\ref{44396100}) without current.
Using (\ref{38920527}), (\ref{40948743}), (\ref{40791520}) and extracting coefficients of the elementary cylindrical
waves $\cylaf{j}{m}\, \e{\bhim\,(k_3\,x^3 - \omega\,x^0)}$ and
 $\cylaf{j}{m}\, \e{\bhim\,(k_3\,x^3 - \omega\,x^0)}$,
we can obtain a system of equations for hyperscalar radial functions $\Fourier{\bDB_{j}^{m}}$ and $\Fourier{\bEH_{j}^{m}}$.
Then simplifying and introducing new unknown functions
\begin{align}
\nonumber
 \Fourier{\bDB_{+}^{m}} &\doteqdot \Fourier{\bDB_{1}^{m}} + \Fourier{\bDB_{-1}^{m}}
\;\;,\quad
&\Fourier{\bEH_{+}^{m}} &\doteqdot \Fourier{\bEH_{1}^{m}} + \Fourier{\bEH_{-1}^{m}}
\;\;,
\\
\label{42303880}
\Fourier{\bDB_{-}^{m}} &\doteqdot \Fourier{\bDB_{1}^{m}} - \Fourier{\bDB_{-1}^{m}}
\;\;,\quad
&\Fourier{\bEH_{-}^{m}} &\doteqdot \Fourier{\bEH_{1}^{m}} - \Fourier{\bEH_{-1}^{m}}
\;\;,
\end{align}
we obtain the following system of equations:
\begin{subequations}
\label{42346412}
\begin{align}
\label{42349616}
{\Fourier{\bEH_{0}^m}}_{;\rho} + \bhim\left(\omega\,\Fourier{\bDB_{-}^{m}} - k_3\,\Fourier{\bEH_{+}^{m}}\right) &= 0
\;\;,\\
\label{42352280}
(\rho\,\Fourier{\bDB_{+}^{m}})_{;\rho} + \bhim\,k_3\,\rho\,\Fourier{\bDB_{0}^{m}} - m\,\Fourier{\bDB_{-}^{m}} &= 0
\;\;,\\
\label{42354864}
(\rho\,\Fourier{\bEH_{-}^{m}})_{;\rho} + \bhim\,\omega\,\rho\,\Fourier{\bDB_{0}^{m}} - m\,\Fourier{\bEH_{+}^{m}} &= 0
\;\;,\\
\label{42358219}
\bhim\rho\left(\omega\,\Fourier{\bDB_{+}^{m}} - k_3\,\Fourier{\bEH_{-}^{m}}\right) +m\,\Fourier{\bEH_{0}^{m}} &= 0
\;\;.
\end{align}
\end{subequations}
where $(...)_{;\rho}$ is the derivative on cylindrical radius $\rho$.

Let us find solutions to system (\ref{42346412}) for the case
$\bDB = \bEH$.
At first, we differentiate equation (\ref{42349616}) by $\rho$, substitute expressions for derivatives
 ${\Fourier{\bDB_{+}^m}}_{;\rho}$ and ${\Fourier{\bDB_{-}^m}}_{;\rho}$ obtained from
(\ref{42352280}) and (\ref{42354864}), and multiply the result by $\rho$. Then we subtract equation $[m\cdot$(\ref{42358219})$/\rho]$
from obtained one and add equation (\ref{42349616}). After multiplying by $\rho$ we obtain as result the equation which contains
the $\Fourier{\bDB_{0}^m}$ component only:
\begin{equation}
\label{39622247}
\left[\rho^2\,\frac{\df^2}{\df \rho^2} + \rho\,\frac{\df}{\df \rho} +
k^2_\rho\,\rho^2 - m^2
\right]\Fourier{\bDB_{0}^m} = 0
\;\;,
\end{equation}
where $k^2_\rho\doteqdot \omega^2 - k_3^2$.

As we see, for the case $k_\rho\neq 0$ equation (\ref{39622247}) corresponds to Bessel equation.
We will use Hankel functions $H^{(1)}_{m}$ and $H^{(2)}_{m}$ (the designations correspond to handbook \cite{AbramStig1964})
for representation its solution. We have
$H^{(2)}_{m}(z) = -\e{\ii\,m\,\pi}\,H^{(1)}_{m}(-z)$. Thus we can use one function $H^{(1)}_{m}$ but with the argument
taking as positive as negative values. Also we must substitute hyperimaginary unit $\bhim$ for imaginary one.
Let us designate this function by the symbol $\RadfunC{m}{}$.
According to \cite{AbramStig1964} we have that $\hconj{\RadfunC{m}{}}(\hconj{z})$ corresponds to Hankel function $H^{(2)}_{m}(z)$.
Thus for real argument $z$, the hyperreal part
$\bRe\RadfunC{m}{}$ is the Bessel function of the first kind and the hyperimaginary part $\bIm\RadfunC{m}{}$ is
the Bessel function of the second kind.

It is convenient to introduce also the following
\begin{Def}
\label{74995577}
The radial cylindrical functions are called the functions
\begin{equation}
\label{75002577}
\RadfunC{m}{k_\rho} \doteqdot k_\rho^{m}\,\RadfunC{m}{}(k_\rho\,\rho)
\;\;,
\end{equation}
where $\RadfunC{m}{}(z)$ is the first Hankel function with hyperimaginary unit instead of imaginary
one, $\rho$ is real and $\rho >0$.
\end{Def}

Using an asymptotic form for the Hankel functions as $z\to 0$ \cite{AbramStig1964}
we can write
\begin{equation}
\label{34826872}
\RadfunC{m}{0} =
-\bhim\,\frac{2^{m}\,\Gamma(m)}{\pi \,{\rho}^{m}}
\qquad \text{for}\qquad m> 0
\;\;,
\end{equation}
where $\Gamma(m)$ is Euler gamma-function. Here expression (\ref{75002577}) is understood in the sense of the
limit $k_\rho\to 0$ for fixed $m$.

Thus the radial cylindrical functions give also finite at infinity ($\rho\to \infty$) solutions to equation (\ref{39622247})
for $k_\rho = 0$ in form (\ref{34826872}).

Taking into account $\hconj{\RadfunC{m}{} (z)} = -(-1)^{m}\,\RadfunC{m}{} (-\hconj{z})$ and using (\ref{75002577}) we obtain
\begin{equation}
\label{67221926}
\hconj{\RadfunC{m}{k_\rho}} = -\RadfunC{m}{-\hconj{k_\rho}}
\;\;.
\end{equation}

Using recurrent relations for the Bessel functions \cite{AbramStig1964} we obtain the following relations for
introduced radial cylindrical functions:
\begin{subequations}
\label{53450696}
\begin{align}
\label{67800509}
&\frac{\df}{\df \rho}\,\RadfunC{m}{k_\rho} = k_\rho^2\,\RadfunC{m-1}{k_\rho} - \frac{m}{\rho}\,\RadfunC{m}{k_\rho}
= -\RadfunC{m+1}{k_\rho} + \frac{m}{\rho}\,\RadfunC{m}{k_\rho}
\;\;,
\\
\label{67803689}
&k^2_\rho\,\RadfunC{m-1}{k_\rho} + \RadfunC{m+1}{k_\rho} = \frac{2\,m}{\rho}\,\RadfunC{m}{k_\rho}
\;\;.
\end{align}
\end{subequations}

According to asymptotic form for the Hankel functions as $|z|\to \infty$ \cite{AbramStig1964}
we have the following asymptotic form for the radial cylindrical functions as $|k_\rho\,\rho|\to\infty$:
\begin{equation}
\label{62616025}
\RadfunC{m}{k_\rho} \sim k_\rho^{m-\frac{1}{2}}\,\sqrt{\frac{2}{\pi\,\rho}}\,
\exp{\left[\bhim\left(k_\rho\,\rho-\frac{m\,\pi}{2}-\frac{\pi}{4}\right)\right]}
\;\;.
\end{equation}

Using the introduced radial cylindrical functions we can write the following form of solution to equation (\ref{39622247}):
\begin{subequations}
\label{45431621}
\begin{equation}
\label{52801349}
\Fourier{\bDB_{0}^m} = C_{k_\rho}^m\,k_{\rho}^2\,\RadfunC{|m|}{k_{\rho}}
\;\;,
\end{equation}
where $C_{k_\rho}^m$ are arbitrary hyperscalar constants,
the factor $k_{\rho}^2$ is introduced for (partial) inclusion the case $k_{\rho}=0$ in
the representation of solution to system (\ref{42346412}).

Then for the case $\omega^2\neq k_3^2$ (i.e. $k_\rho^2\neq 0$) we can obtain directly solutions for
$\Fourier{\bDB_{+}^m}$ and $\Fourier{\bDB_{-}^m}$ from
system of equations (\ref{42349616}) and (\ref{42358219}) ($\Fourier{\bDB_{j}^m} = \Fourier{\bEH_{j}^m}$).
By direct substitution we have that the obtained solution for
$\Fourier{\bDB_{+}^m}$ and $\Fourier{\bDB_{-}^m}$ satisfies also equations (\ref{42352280}) and (\ref{42354864}).
Using relations (\ref{42303880}) and (\ref{67800509}) we have
\begin{equation}
\label{52893372}
\Fourier{\bDB_{\pm 1}^m} = \frac{C_{k_\rho}^m\,\bhim}{2}\,\left(\omega\pm k_3\right)
\left(\frac{m\pm |m|}{\rho}\,\RadfunC{|m|}{k_\rho}\mp \RadfunC{|m|+1}{k_\rho}\right)
\;,
\end{equation}
where $k_\rho = \pm\sqrt{\omega^2 - k_3^2}$ for $\omega^2 > k_3^2$ and
$k_\rho = \pm\bhim\,\sqrt{k_3^2 - \omega^2}$ for $\omega^2 < k_3^2$.
\end{subequations}

Let us introduce the following designation:
\begin{equation}
\label{53130640}
\garmcb{\omega}{k_3}{\pm}{m} \doteqdot
 \left(\Fourier{\bDB_{0}^m}\,\cylaf{0}{m} +  \Fourier{\bDB_{1}^m}\,\cylaf{1}{m} +
 \Fourier{\bDB_{-1}^m}\,\cylaf{-1}{m}\right)\e{\bhim\,k_3\,x^3}
\;\;,
\end{equation}
where $\Fourier{\bDB_{j}^m}$ are taken from (\ref{45431621}) with substitution $C_{k_\rho}^m=1$,
left-hand index $+$ or $-$ corresponds to one case from the two $k_\rho = \pm\sqrt{\omega^2 - k_3^2}$
(the both cases $\omega^2\gtrless k_3^2$
are considered, $\sqrt{-1}\doteqdot \bhim$).

Thus an elementary cylindrical solution to equation (\ref{44396100}) without sources and for the case $\bDB=\bEH$
can be written in the form
\begin{equation}
\label{76377847}
\garmcb{\omega}{k_3}{\pm}{m}\, \e{-\bhim\, \omega\,x^0}
\;\;.
\end{equation}

It is evident that $\garmcb{\omega}{k_3}{\pm}{m}$ are eigenfunctions of self-adjoint operator
$-\bhim\,\bnp$:
\begin{equation}
\label{56023850}
-\bhim\,\bnp\,\garmcb{\omega}{k_3}{\pm}{m} = \omega\,\garmcb{\omega}{k_3}{\pm}{m}
\;\;.
\end{equation}

According to asymptotic (\ref{62616025}) we have that the case of
positive real $k_\rho = \sqrt{\omega^2 - k_3^2}$
(function $\garmcb{\omega}{k_3}{+}{m}$ into (\ref{76377847})) corresponds to
 a divergent radial wave (from $x_3$ axis in addition to a propagation along it)
  and  the case of negative real $k_\rho = -\sqrt{\omega^2 - k_3^2}$
(function $\garmcb{\omega}{k_3}{-}{m}$ into (\ref{76377847}))
corresponds to the convergent radial wave.

In general case both the divergent and convergent waves
must be considered for each time harmonic $\e{-\bhim\,\omega\,x^0}$
(two harmonics differing with a sign of cyclic frequency $\omega$
are considered as different).
If we change $\RadfunC{|m|}{k_\rho}$ to $\bRe\RadfunC{|m|}{k_\rho}$ into (\ref{45431621})
then we have an everywhere regular solution for real $k_\rho$.
Thus in view of (\ref{67221926}) we can write the following everywhere regular and finite
(for the case $\omega^2 > k_3^2$) functions:
\begin{equation}
\label{42066794}
\garmcb{\omega}{k_3}{\bumpeq}{m} \doteqdot \frac{1}{2}\left(\garmcb{\omega}{k_3}{+}{m} -\garmcb{\omega}{k_3}{-}{m}\right)
\;\;.
\end{equation}
The appropriate elementary solutions
\begin{equation}
\label{42527967}
\garmcb{\omega}{k_3}{\bumpeq}{m} \e{-\bhim\, \omega\,x^0}
\end{equation}
are radial-undistorted waves propagating along the $x^3$ axis.
For the case of $\omega^2 > k_3^2$ these solutions have the form of standing waves in $(\rho,\varphi)$ plane
and also are called Bessel beams. A phase velocity of these beams
exceeds the velocity of light ($1< |\omega/k_3| < \infty$) but their group velocity less then the velocity of light
($|\p \omega/\p k_3|<1$).

The case of hyperimaginary  $k_\rho = \pm\bhim\,\sqrt{k_3^2 - \omega^2}$ ($\omega^2 < k_3^2$) also must be considered.
This case is connected with Hankel functions of imaginary argument. The appropriate radius-infinity
asymptotic is obtained from (\ref{62616025}). This case can be interested in nonlinear theory and for problems with
boundaries such that in waveguide. A phase velocity along $x^3$ axis of the solutions with $\omega^2 < k_3^2$
less then the velocity of light ($|\omega/k_3|<1$)
but their group velocity exceeds the velocity of light ($1< |\p \omega/\p k_3|< \infty$). Solutions
with these unusual properties are known \cite{ZamboniRached2002a}.

For the cases $\omega = \pm k_3 \neq 0$, a combination of equations (\ref{42349616}) and (\ref{42358219}) gives two first order equations
for $\Fourier{\bDB_{0}^m}$. Then we solve them and substitute obtained expressions for $\Fourier{\bDB_{0}^m}$ into equations
(\ref{42352280}) and (\ref{42354864}). We obtain solutions to these equations,
the cases $m=\omega/k_3$ are considered separately. As result we can write the following solutions:
\begin{subequations}
\label{47363782}
\begin{align}
\nonumber
 & \omega = k_3\;\;\Longrightarrow  \quad\quad\Fourier{\bDB_{0}^m} =
 \frac{{}_{+}\!C^{m}_{\mspace{-5mu}\prime\prime 0}}{\rho^{m}}\;\;,
\\
\label{47374007}
&\quad\quad\Fourier{\bDB_{1}^m} = \frac{{}_{+}\!C^{m}_{\mspace{-5mu}\prime\prime 0}\,\bhim\,\omega}{2\left(m-1\right)\rho^{m-1}} +
\frac{{}_{+}\!C^{m}_{\mspace{-3mu}\prime\mspace{1mu} 0}}{\rho^{1-m}}
\;\;,\qquad
 \Fourier{\bDB_{-1}^m} = \frac{{}_{+}\!C^{m}_{\mspace{-5mu}\prime\prime 0}\,\bhim\,m}{2\,\omega\, \rho^{m+1}}
\;\;;\\[1.5ex]
\nonumber
 & \omega = -k_3\;\;\Longrightarrow  \quad\quad\Fourier{\bDB_{0}^m} = \frac{{}_{-}\!C^{m}_{\mspace{-5mu}\prime\prime 0}}{\rho^{-m}}\;\;,
\\
\label{48830231}
&\quad\quad  \Fourier{\bDB_{1}^m} = \frac{{}_{-}\!C^{m}_{\mspace{-5mu}\prime\prime 0}\,\bhim\,m}{2\,\omega\, \rho^{1-m}}
\;\;,\qquad
\Fourier{\bDB_{-1}^m} = \frac{{}_{-}\!C^{m}_{\mspace{-5mu}\prime\prime 0}\,\bhim\,\omega}{2\left(m+1\right)\rho^{-m-1}} +
\frac{{}_{-}\!C^{m}_{\mspace{-3mu}\prime\mspace{1mu} 0}}{\rho^{m+1}}
\;\;.
\end{align}
\end{subequations}
Here for the special cases $m=\omega/k_3=\pm 1$
 we must take ${}_{\pm}\!C^{m}_{\mspace{-5mu}\prime\prime 0}=0$.
There is also a logarithmic solution (containing the term $\log\rho$) for these cases but it is not considered here.

Let us consider solutions (\ref{47363782}) that are finite as $\rho\to \infty$. These solutions are infinite at $\rho=0$.
But they can be interested in nonlinear theory for asymptotic behaviour of solution as $\rho\to \infty$.
We have these solutions if we take into (\ref{47363782}) the following:
\begin{equation}
\label{57559679}
{}_{\pm}\!C^{m}_{\mspace{-5mu}\prime\prime 0} = 0\quad\text{for}\quad \frac{\omega}{k_3}\,m \leqslant 1
\;\;,\quad\qquad
{}_{\pm}\!C^{m}_{\mspace{-3mu}\prime\mspace{1mu} 0} = 0\quad\text{for}\quad \frac{\omega}{k_3}\,m \geqslant 2
\;\;,
\end{equation}
where $\omega\neq 0$. For the case
$m\,{\omega}/{k_3} < 0$
(${}_{\pm}\!C^{m}_{\mspace{-5mu}\prime\prime 0} = 0$)
solution (\ref{47363782}) is given by formulas (\ref{45431621}) with $\omega = \pm k_3$. A connection between
the free constants ${}_{\pm}\!C^{m}_{\mspace{-3mu}\prime\mspace{1mu} 0}$
in (\ref{47363782}) and $C^{m}_{0}$ in (\ref{45431621})
 can be obtained with the help of
(\ref{34826872}). The case $m\,{\omega}/{k_3} \geqslant 2$ (${}_{\pm}\!C^{m}_{\mspace{-3mu}\prime\mspace{1mu} 0} = 0$)
is not described by formula (\ref{45431621}).

For the case $\omega = k_3 = 0$ we must take the sum of two solutions (\ref{47374007}) and (\ref{48830231}) with
${}_{\pm}\!C^{m}_{\mspace{-5mu}\prime\prime 0} = 0$
for $m\neq 0$. If we take
${}_{+}\!C^{m}_{\mspace{-3mu}\prime\mspace{1mu} 0}=0$ for $m\geqslant 2$ and
${}_{-}\!C^{m}_{\mspace{-3mu}\prime\mspace{1mu} 0}=0$ for $m\leqslant -2$ the solution is finite at $\rho$-infinity.

As we can see, for the case $m\,{\omega}/{k_3} = 1$ ($\omega\neq 0$) in (\ref{47363782}) we have the following two
solutions to equation (\ref{48931593}) in the form of plane waves with constant amplitudes:
\begin{subequations}
\label{55849795}
\begin{align}
\label{55859098}
& \cylaf{1}{1}\, \e{\bhim\,(k_3\,x^3 - \omega\,x^0)}
&\text{for}
\qquad
\omega &= k_3
\;\;,\\
\label{55867560}
& \cylaf{-1}{-1}\, \e{\bhim\,(k_3\,x^3 - \omega\,x^0)}
&\text{for}
\qquad
\omega &= -k_3
\;\;.
\end{align}
\end{subequations}
In view of (\ref{12342156}) and (\ref{qqqwddtyh}) we conclude that
for $\omega > 0$ solution (\ref{55859098}) is the clockwise polarized mode propagating in positive direction
of $x^3$ axis but for $\omega < 0 $ it is the counterclockwise polarized mode. And
solution (\ref{55867560}) for $\omega > 0$ is the clockwise polarized mode propagating in negative direction
of $x^3$ axis but for $\omega < 0 $ it is the counterclockwise polarized mode.
Thus the sign of circular frequency $\omega$ is the sign of rotation for circularly polarized waves but the indices of cylindrical angle functions
connected with a direction of propagation for this case.

A plane wave of arbitrary polarization is represented with the sum of positive and negative frequency components
(for (\ref{55859098}) or (\ref{55867560}) waves) with arbitrary quasi-hyperscalar coefficients.
These two coefficients (four real numbers) define a po\-la\-ri\-za\-tion ellipse and an amplitude of wave.

\section{Spherical waves}
\label{sphericalwaves}
Now let us consider a spherical coordinate system $\{x^0,\,r,\,\vartheta,\,\varphi\}$ in flat space-time
($x^0 = -x_0$, $x^i = x_i$):
$x_1 = r\,\sin\vartheta\,\cos\varphi$,
$x_2 = r\,\sin\vartheta\,\sin\varphi$,
$x_3 = r\,\cos\vartheta$,
where $0\leqslant\vartheta\leqslant\pi$ and $0\leqslant\varphi < 2\,\pi$.
Let us call $\vartheta$ vertical angle (counting off from vertical line) and $\varphi$ horizontal angle.

\begin{subequations}
\label{48706537}
\begin{Def}
\label{48282627}
The angle spherical functions are called the
even space-time hypernumber
eigenfunctions depending on
vertical $\vartheta$ and horizontal  $\varphi$ spherical angles
for the operators $\bRotS_{3}$ and $(\bRotS)^2$, and are denoted by $\spheraf{j}{l}{m}$:
\begin{align}
\label{48779195}
\bRotS_3\,\spheraf{j}{l}{m} &= m\,\spheraf{j}{l}{m}
\;\;,\\
\label{41272787}
(\bRotS)^2\,\spheraf{j}{l}{m} &= l\left(l+1\right)\spheraf{j}{l}{m}
\;\;,
\end{align}
where $\spheraf{j}{l}{m}= \spheraf{j}{l}{m}(\vartheta,\varphi)$.
\end{Def}
\end{subequations}

We take $j=\mathrm{s}$ for hyperscalar and $j=0,\pm 1$ for bivector
functions.

Raising and reducing operators (\ref{wwsderggv}) have the following form in spherical coordinates:
\begin{subequations}
\label{35537767}
\begin{align}
\label{35545546}
\bRotS_{+} &= \e{\bhim\,\varphi}\left(\frac{\p}{\p \vartheta} + \bhim\,\cot\vartheta\,\frac{\p}{\p \varphi}
+ \cylaf{1}{0}\bwedge\right)
\;\;,\\
\label{35551326}
\bRotS_{-} &= \e{-\bhim\,\varphi}\left(-\frac{\p}{\p \vartheta} + \bhim\,\cot\vartheta\,\frac{\p}{\p \varphi}
+ \cylaf{-1}{0}\bwedge\right)
\;\;.
\end{align}
\end{subequations}

From (\ref{40047342}), (\ref{35537767}), and (\ref{ddlfkgjnneexs}) it follows:
\begin{subequations}
\label{60946513}
\begin{align}
\label{61072414}
(\bRotS)^{2} &= -\frac{\p^2}{\p \vartheta^2} - \cot\vartheta\,\frac{\p}{\p \vartheta} -\frac{1}{\sin^2\vartheta}\,\frac{\p^2}{\p \varphi^2}
\\
\label{61075638}
 +& \left(\cylaf{-1}{0}-\cylaf{1}{0}\right)\bwedge\frac{\p}{\p \vartheta} +
\bhim\left[\cot\vartheta\left(\cylaf{1}{0}+\cylaf{-1}{0}\right) - 2\,\cylaf{0}{0}\right]\bwedge\frac{\p}{\p \varphi} + 2
\;\;.
\end{align}
\end{subequations}
Here the formula is divided for two parts (\ref{61072414}) and (\ref{61075638}).
The first part act to hyperscalar and bivector functions but the second part act only to bivector functions.
The second part contains a multiplication by $2$. There is not this operation for the case of hyperscalar function.

By analogy with (\ref{wlllrokjds}) we represent the operator of space differentiation $\bnp$ (\ref{wkdhjfof})
in spherical coordinates:
\begin{equation}
\label{79055330}
\bnp = \bbaab^{r}\,\frac{\p}{\p r} + \bbaab^{\vartheta}\,\frac{\p}{\p \vartheta} + \bbaab^{\varphi}\,\frac{\p}{\p \varphi}
\doteqdot \bnp^{r} + \bnp^{\vartheta} + \bnp^{\varphi}
\;\;,
\end{equation}
where $\bbaab^{r}\doteqdot \baab^{r}\bwedge\baab^0 = \bbaab_{r}$, $\bbaab^{\vartheta}\doteqdot \baab^{\vartheta}\bwedge\baab^0$.
The metric tensor for spherical coordinate system ($\metr_{rr} = 1$, $\metr_{\vartheta\vartheta} = r^2$,
$\metr_{\varphi\varphi}=\rho^2 = r^2\,\sin^2\vartheta$) gives relations $\bbaab_{\vartheta} = r^2\,\bbaab^{\vartheta}$,
$\bbaab_{\varphi} = r^2\,\sin^2\vartheta\,\bbaab^{\varphi}$.

We have
\begin{subequations}
\label{38955361}
\begin{align}
\label{38964023}
 & \bbaab_{r} = \bbaab_{\rho}\,\sin\vartheta +\bbaab_{3}\,\cos\vartheta
= \frac{1}{2}\left(\cylaf{-1}{0} + \cylaf{1}{0}\right)\sin\vartheta + \cylaf{0}{0}\,\cos\vartheta
\;,
\\
\label{38967508}
 & \bbaab^{\vartheta} = \frac{1}{r}\left(\bbaab_{\rho}\,\cos\vartheta - \bbaab_{3}\,\sin\vartheta\right) =
\frac{1}{r}\left[\frac{1}{2}\left(\cylaf{-1}{0} + \cylaf{1}{0}\right)\cos\vartheta -
\cylaf{0}{0}\,\sin\vartheta\right]
\;,
\\
\label{38040525}
 & \bbaab^{\varphi} = \frac{\bhim}{2\,r\,\sin\vartheta}\left(\cylaf{-1}{0} - \cylaf{1}{0}\right)
\;.
\end{align}
\end{subequations}

Using (\ref{38955361}) we can rewrite operator $(\bRotS)^{2}$ (\ref{60946513}) in the form:
\begin{subequations}
\label{62841618}
\begin{align}
\label{62845814}
(\bRotS)^{2} &= -\frac{1}{\sin\vartheta}\,\frac{\p}{\p \vartheta}\,\sin\vartheta\,\frac{\p}{\p \vartheta}
 -\frac{1}{\sin^2\vartheta}\,\frac{\p^2}{\p \varphi^2}
\\
\label{62872162}
 &\quad\quad - 2\,\bhim\,r\,\sin\vartheta\left[\bbaab^{\varphi}\bwedge\frac{\p}{\p \vartheta} -
\bbaab^{\vartheta}\bwedge\frac{\p}{\p \varphi}\right] + 2
\;\;.
\end{align}
\end{subequations}

Existing results relating to rotation group (see \cite{GelfMilnShap1958e,Vilenkin1968}) give guide messages which allow
to consider the spherical angle functions in the following form:
\begin{equation}
\label{47394428}
\spheraf{j}{l}{m} = \spherzonf{j}{l}{m} \,\sphersecf{j}{m}
\qquad\left(\nosum\right)\;\;,
\end{equation}
where $\spherzonf{j}{l}{m} = \spherzonf{j}{l}{m}(\zc)$, $\zc\doteqdot\cos\vartheta$ but in general case
$\sphersecf{j}{m} = \sphersecf{j}{m} (\vartheta,\varphi)$,
$l$ will be called zonal index and  $m$ sectorial one, $-l\leqslant m \leqslant l$.

For the case of hyperscalar spherical function we have that $\spherzonf{\mathrm{s}}{l}{0}$ is the \mbox{l-th}
Legendre polynomial
and $\spherzonf{\mathrm{s}}{l}{m}$ are connected with the associated Legendre polynomials.
I call the functions $\spherzonf{j}{l}{m}$ zonal spherical functions or harmonics. This appellation looks justified because
these functions divide a sphere into zones with the borders in the form of equator parallel circles\footnote%
{The appellation ``zonal harmonics'' occurs also in literature for scalar angle spherical functions with $m=0$. But here we will use this
appellation for the functions $\spherzonf{j}{l}{m}(\zc)$.}.
Within a zone the sign of each multivector component of function $\spherzonf{j}{l}{m}$ is invariable.
We consider quasi-hyperscalar zonal spherical harmonics.

For the case of hyperscalar angle spherical functions we have the following
sectorial harmonics\footnote%
{The appellation ``sectorial harmonics'' occurs also in literature for scalar angle spherical functions with $m=l$. But here we will use this
appellation for the functions $\sphersecf{j}{m}$}:
\begin{subequations}
\label{37994773}
\begin{equation}
\label{38004600}
\sphersecf{\mathrm{s}}{m}= \e{\bhim\,m\,\varphi} \doteqdot \cylaf{\mathrm{s}}{m}
\;\;.
\end{equation}
Sectorial harmonics divide a sphere into zones with the borders in the form of  circles passing through poles.
Within a sector the sign of each multivector component of function $\sphersecf{j}{m}$ is invariable.

Sectorial bivector function depends on both vertical and horizontal angles:
$\sphersecf{j}{m} = \sphersecf{j}{m} (\vartheta,\varphi)$, $j=0,\pm 1$.
It is convenient to use the following sectorial spherical harmonics:
\label{48024480}
\begin{align}
\label{48033624}
\sphersecf{0}{m} &\doteqdot \bbaab_{r}\,\e{\bhim\,m\,\varphi}
= \frac{1}{2}\left(\cylaf{-1}{m} + \cylaf{1}{m}\right)\sin\vartheta + \cylaf{0}{m}\,\cos\vartheta
\;,
\\
\nonumber
\sphersecf{\pm 1}{m} &\doteqdot \left(r\,\bbaab^{\vartheta} \pm r\,\sin\vartheta\,\bhim\,\bbaab^{\varphi}\right)\e{\bhim\,m\,\varphi}
\\
\label{48036593}
&\qquad\qquad =
\cylaf{\pm 1}{m}\,\cos^2\frac{\vartheta}{2} - \cylaf{\mp 1}{m}\,\sin^2\frac{\vartheta}{2} - \cylaf{0}{m}\,\sin\vartheta
\;\;.
\end{align}
\end{subequations}
Accordingly, we have the following expression for basis bivectors of spherical coordinate system:
\begin{equation}
\label{33776714}
\bbaab_{r} = \sphersecf{0}{0}
\;\;,
\quad
\bbaab^{\vartheta} = \frac{1}{2\,r}\left(\sphersecf{1}{0} + \sphersecf{-1}{0}\right)
\;\;,
\quad
\bbaab^{\varphi} = \frac{\bhim}{2\,r\,\sin\vartheta}\left(\sphersecf{-1}{0} - \sphersecf{1}{0}\right)
\;\;.
\end{equation}

Connection between cylindrical and spherical angle functions (\ref{37994773}) can be also represented as
a local space rotation through angle $\vartheta$ about the unit bivector $r\,\sin\vartheta\,\bbaab^\varphi$.
In general, space-time rotation for multivectors in space-time Clifford algebra is realized by means of some even hypernumber
$\bHCLor$ such that
$\mathbf{C}^{\prime} = \bHCLor\,\mathbf{C}\,\invop{\bHCLor}$ (see, for example, \cite{HestenesSobczyk1984}).
Thus we have the following representation for expressions (\ref{37994773}):
\begin{equation}
\label{62941079}
\sphersecf{j}{m} = \bHCLorS_{S}\,\cylaf{j}{m}\,\invop{\bHCLorS_{S}}
\doteqdot \exp\biggl(-\bhim\,\bar{\bbaab}^\varphi\,\frac{\vartheta}{2}\biggr)\cylaf{j}{m}
\exp\biggl(\bhim\,\bar{\bbaab}^\varphi\,\frac{\vartheta}{2}\biggr)
\;\;,
\end{equation}
where $j=\mathrm{s},0,1,-1$, $\bar{\bbaab}^\varphi\doteqdot\bbaab^\varphi\,r\,\sin\vartheta$,
\begin{align}
\nonumber
&\bHCLorS_{S}^{\pm 1} = \exp\left[\frac{1}{4}\left(\cylaf{\mp 1}{0}-\cylaf{\pm 1}{0}\right)\vartheta\right]
= \exp\left[\frac{1}{4}\left(\sphersecf{\mp 1}{0}-\sphersecf{\pm 1}{0}\right)\vartheta\right]
\\
\label{44204762}
 &=  \cos\frac{\vartheta}{2} + \frac{1}{2}\left(\cylaf{\mp 1}{0}-\cylaf{\pm 1}{0}\right)\sin\frac{\vartheta}{2}
= \cos\frac{\vartheta}{2} + \frac{1}{2}\left(\sphersecf{\mp 1}{0}-\sphersecf{\pm 1}{0}\right)\sin\frac{\vartheta}{2}
\;\;.
\end{align}

In view of formula (\ref{62941079}) a multiplication table for the sectorial harmonics has
the same form as for cylindrical angle functions (\ref{40948743}):
\begin{align}
\nonumber
&  \sphersecf{1}{m}\,\sphersecf{1}{n} = \sphersecf{-1}{m}\,\sphersecf{-1}{n} = 0
\;\;,\quad
\sphersecf{1}{m}\bcdot\sphersecf{0}{n} = \sphersecf{-1}{m}\bcdot\sphersecf{0}{n} = 0
\;\;,\\
\nonumber
& \sphersecf{1}{m}\bcdot\sphersecf{-1}{n} = 2\,\sphersecf{\mathrm{s}}{m+n}
\;\;,\quad
\sphersecf{0}{m}\bcdot\sphersecf{0}{n} = \sphersecf{\mathrm{s}}{m+n}
\;\;,\quad
\sphersecf{j}{m}\,\sphersecf{\mathrm{s}}{n} = \sphersecf{\mathrm{s}}{n}\,\sphersecf{j}{m} = \sphersecf{j}{m+n}
\;\;,\\
\nonumber
& \sphersecf{1}{m}\bwedge\sphersecf{-1}{n} = 2\,\sphersecf{0}{m+n}
\;\;,\quad
\sphersecf{1}{m}\bwedge\sphersecf{0}{n} = -\sphersecf{1}{m+n} \;\;,\quad  \sphersecf{-1}{m}\bwedge\sphersecf{0}{n} = \sphersecf{-1}{m+n}
\;\;,
\\
\label{37150223}
&
\hconj{\sphersecf{\mathrm{s}}{m}} = \sphersecf{\mathrm{s}}{-m}
\;\;,\quad
\hconj{\sphersecf{0}{m}} = \sphersecf{0}{-m}
\;\;,\quad
\hconj{\sphersecf{1}{m}} = \sphersecf{-1}{-m}
\;\;,\quad
\hconj{\sphersecf{-1}{m}} = \sphersecf{1}{-m}
\;\;.
\end{align}

Obviously (because of (\ref{37994773})), spherical angle functions (\ref{47394428}) are eigenfunctions with eigenvalue $m$
for operator $\bRotS_3$ (\ref{ddlfkgjnneexs}).

By substituting (\ref{47394428}) into (\ref{41272787}), taking into account (\ref{62841618}) and (\ref{37994773}), and using new variable
$\zc\doteqdot\cos\vartheta$, we obtain the following equation for zonal harmonics:
\begin{equation}
\label{58132433}
\left[\left(1 - \zc^2\right) \frac{\df^2}{\df\zc^2} - 2\,\zc\,\frac{\df}{\df\zc} - \frac{j^2 + m^2 - 2\,m\,j\,\zc}{1 - \zc^2} + l\left(l+1\right)\right]
\spherzonf{j}{l}{m} = 0
\;\;.
\end{equation}
Here for the cases of hyperscalar and bivector spherical functions we have
 $j=\mathrm{s},0,1,-1$ (we take $\mathrm{s}=0$ into operator). Thus $\spherzonf{\mathrm{s}}{l}{m} \doteqdot \spherzonf{0}{l}{m}$
but $\sphersecf{\mathrm{s}}{m}\neq\sphersecf{0}{m}$.

Solutions to equation (\ref{58132433}) are known. The appropriate functions are described by Gelfand
with co-authors \cite{GelfMilnShap1958e}
and investigated in detail by Vilenkin \cite{Vilenkin1968} (where the designation $P^l_{nm}$ is used for these functions).
According to \cite{Vilenkin1968} we have the following formula for zonal harmonics
(the imaginary unit is changed by the hyperimaginary one):
\begin{multline}
\label{34477901}
\spherzonf{j}{l}{m}(\zc) {} = {}
\frac{(-1)^{l-j}\,\bhim^{j-m}}{2^l}
\sqrt{\frac{(l {}+{} m)!}{(l {}-{} j)!\,(l {}+{} j)!\,(l {}-{} m)!}}
\\[11pt]
{}\cdot{} (1 {}+{} \zc)^{-\frac{1}{2}\,(m {}+{} j)}\,(1 {}-{} \zc)^{\frac{1}{2}\,(j-m)}\,
\dfrac{d^{l-m}}{d\zc^{l-m}}
\left[(1-\zc)^{l-j}\,(1 {}+{} \zc)^{l {}+{} j}\right]
\;\;,
\end{multline}
where $l\geqslant 0$, $|j|\leqslant l$, $|m|\leqslant l$. Here we consider the case when all indices
are whole numbers\footnote%
{Formula (\ref{34477901}) gives also finite at the points $\zc=\pm 1$ solutions to equation (\ref{58132433}) for the case when
all indices at functions $\spherzonf{j}{l}{m}$ be half-integer numbers \cite{Vilenkin1968}.
This case is connected with double-valued representations for rotation group.
Even hypernumber functions of type $\bHCLorS_{S}$ (\ref{44204762})
using for a local rotation of multivectors (see (\ref{62941079}))
realizes such representation.
But at the present paper this case is not considered.}.
Thus the minimal value of zonal index $l$ for the zonal harmonics $\spherzonf{\mathtt{s}}{l}{m}$ and $\spherzonf{0}{l}{m}$
is $0$ but for the zonal harmonics $\spherzonf{1}{l}{m}$ and $\spherzonf{-1}{l}{m}$ is $1$.

The functions defined as (\ref{34477901}) have, in particular, the following properties:
\begin{subequations}
\label{41580284}
\begin{equation}
\label{41187219}
\spherzonf{j}{l}{m} = \spherzonf{-j}{l}{,-m}
\;\;,\quad
\hconj{\spherzonf{j}{l}{m}} = (-1)^{j-m}\,\spherzonf{j}{l}{m}
\;\;,\quad
\spherzonf{j}{l}{m} = \spherzonf{m}{l}{j}
\;\;.
\end{equation}
\begin{equation}
\label{43834239}
\int\limits_{-1}^{1}
\spherzonf{j}{l}{m}\,\hconj{\spherzonf{j}{l^{\prime}}{m}}
\,\df \zc =  \frac{2}{2\,l + 1}\,\delta_{l l^{\prime}}
\;\;.
\end{equation}
\end{subequations}
Also we have the recurrent formulas
\begin{equation}
\label{45276893}
\left(\sqrt{1 - \zc^2}\;\frac{\df }{\df\zc} {}\pm{} \frac{j\,\zc -m}{\sqrt{1 - \zc^2}}\right)\spherzonf{j}{l}{m} =
-\bhim\,\sqrt{(l \mp j)\,(l \pm j +1)}\;\spherzonf{j\pm 1}{l}{,m}
\end{equation}
and multiplication table for the zonal harmonics:
\begin{equation}
\label{45992823}
\spherzonf{j_1}{l_1}{m_1}\,\spherzonf{j_2}{l_2}{m_2} = \sum_{l_{\min}}^{l_1 + l_2}
\ClGord{l_1}{l_2}{l}{j_1}{j_2}\,\ClGord{l_1}{l_2}{l}{m_1}{m_2}\,\spherzonf{j_1 + j_2}{l}{,\,m_1+m_2}
\;\;,
\end{equation}
where $l_{\min} = \max (|l_1-l_2|,\,|j_1+j_2|,\,|m_1+m_2|)$
and $\ClGord{l_1}{l_2}{l}{m_1}{m_2}$ are Clebsch-Gordan coefficients for rotation group. According to \cite{Vilenkin1968}
we have the following formula:
\begin{multline}
\label{46589541}
\ClGord{l_1}{l_2}{l}{m_1}{m_2} = \sqrt{\left(2\,l + 1  \right)}
\\
\cdot \sqrt{\frac{\left( {l_1} + {m_1} \right)!  \left( l - {m_1} - {m_2} \right)! \left( l - l_1 + l_2 \right) !
    \left( l_1 + {l_2} -l \right)! \left( {l_1} + {l_2} + l + 1 \right)!}%
{\left( {l_1} - {m_1} \right)! \left( {l_2} + {m_2} \right)! \left( {l_2} - {m_2} \right)!
    \left( l + {m_1} + {m_2} \right)! \left( l + {l_1} - {l_2} \right)!}}
\\
\cdot \sum_{l^{\prime}=l^{\prime}_{\text{min}}}^{l}
\frac{{\left( -1 \right) }^{{l_1} + {m_2} - l^{\prime}} \left( l + l^{\prime} \right) !
    \left( {l_2} + l^{\prime} - {m_1} \right) !}{ \left( l - l^{\prime} \right)! \left(l^{\prime} -{m_1} - {m_2} \right)!
    \left(l^{\prime}  -{l_1} + {l_2} \right)! \left({l_1} + {l_2} + l^{\prime} + 1 \right)! }
\;\;,
\end{multline}
where
$l^{\prime}_{\text{min}} = \max ({m_1} + {m_2}, {l_1} - {l_2})$.

According to (\ref{47394428}), (\ref{37150223}), and (\ref{41187219}) we have the following rule of hyperconjugation
for the angle spherical functions:
\begin{equation}
\label{55376750}
\hconj{\spheraf{j}{l}{m}} = (-1)^{j-m}\,\spheraf{-j}{l}{,-m}
\;\;.
\end{equation}

According to  (\ref{37150223}) and (\ref{43834239}) we have relation of orthogonality on sphere for the angle spherical functions
\begin{multline}
\label{32710392}
\prodf{\spheraf{j}{l}{m}}{\spheraf{j^{\prime}}{l^{\prime}}{m^{\prime}}}_{\vartheta\varphi}
\doteqdot
\int\limits_{0}^{2\,\pi} \df\varphi\int\limits_{0}^{\pi} \spheraf{j}{l}{m}\bcdot\hconj{\spheraf{j^{\prime}}{l^{\prime}}{m^{\prime}}}
\,\sin\vartheta\,\df\vartheta
\\
= \frac{4\,\pi\left(1+j^2\right)}{2\,l + 1}\,\delta_{l l^{\prime}}\,\,\delta_{j j^{\prime}}\,\delta_{m m^{\prime}}
\;\;,
\end{multline}
where $j=\mathrm{s},1,0,-1$ and we take $\mathrm{s}=0$ into right-hand part.

Expansion of an even hypernumber function in the spherical angle harmonics has the following form:
\begin{equation}
\label{49438148}
\mathbf{Q} (x^0,r,\vartheta,\varphi) = \sum_{j}^{\mathrm{s},0,1,-1}\,\sum_{l=|j|}^{\infty}\,\sum_{m=-l}^{l}\spheraf{j}{l}{m}\,
\int\limits_{-\infty}^{+\infty} \Fourier{\mathbf{Q}_{j}^{lm}}\,\e{-\bhim\,\omega\,x^0}\,\df \omega
\;\;,
\end{equation}
where for $j = \mathrm{s}$ we take $|\mathrm{s}|\doteqdot 0$,
$\Fourier{\mathbf{Q}_{j}^{lm}}$ are quasi-hyperscalars:
\begin{equation}
\label{71561089}
\Fourier{\mathbf{Q}_{j}^{lm}}(r,\,\omega) =
\frac{1}{2\pi\,\norma{\spheraf{j}{l}{m}}^2}\,\prodf{\mathbf{Q} (x^0,r,\vartheta,\varphi)}%
{\spheraf{j}{l}{m}\,\e{\bhim\,\omega\,x^0}}_{\vartheta\varphi x^{0}}
\;\;,
\end{equation}
where $\norma{\spheraf{j}{l}{m}}^2 = [4\,\pi\,(1+j^2)]/(2\,l + 1)$ according to (\ref{32710392}).

Using (\ref{33776714}), (\ref{47394428}), (\ref{37150223}), and (\ref{45276893}) we obtain the following table:
\begin{align}
\nonumber
&r\,\bnpc\bcdot \spheraf{\mathrm{s}}{l}{m} = \frac{\bhim}{2}\,\sqrt{l\,(l+1)}\left(\spheraf{1}{l}{m}+\spheraf{-1}{l}{m}\right)
\;\;,\qquad
r\,\bnpc\bwedge \spheraf{\mathrm{s}}{l}{m} = 0
\;\;,\\
\nonumber
&r\,\bnpc\bcdot \spheraf{0}{l}{m} = 2\,\spheraf{\mathrm{s}}{l}{m}
\;\;,\qquad
r\,\bnpc\bwedge \spheraf{0}{l}{m} = \frac{\bhim}{2}\,\sqrt{l\,(l+1)}\left(\spheraf{-1}{l}{m}-\spheraf{1}{l}{m}\right)
\;\;,\\
\nonumber
&r\,\bnpc\bcdot \spheraf{\pm 1}{l}{m}  =  \bhim\,\sqrt{l\,(l+1)}\,\spheraf{\mathrm{s}}{l}{m}
\;\;,\quad
\\
\label{49414433}
&\qquad\qquad\qquad\qquad\qquad
r\,\bnpc\bwedge \spheraf{\pm 1}{l}{m} = \pm \bigl[\spheraf{\pm 1}{l}{m} - \bhim\,\sqrt{l\,(l+1)}\,\spheraf{0}{l}{m}\bigr]
\;\;,
\end{align}
where $\bnpc\doteqdot\bnp^{\vartheta} + \bnp^{\varphi}$ is the angle part of space differentiation operator (\ref{79055330}).

We have also the table for multiplication of the radial basis bivector by the angle spherical functions:
\begin{align}
\nonumber
& \bbaab_r\bcdot \spheraf{\mathrm{s}}{l}{m} = \spheraf{0}{l}{m}
\;\;,\quad
\bbaab_r\bcdot \spheraf{0}{l}{m} = \spheraf{\mathrm{s}}{l}{m}
\;\;,\quad
\bbaab_r\bwedge \spheraf{\mathrm{s}}{l}{m} = \bbaab_r\bwedge \spheraf{0}{l}{m} = 0
\;\;,\\
\label{47086966}
&\bbaab_r\bcdot \spheraf{\pm 1}{l}{m}  = 0
\;\;,\qquad
\bbaab_r\bwedge \spheraf{\pm 1}{l}{m} = \pm\spheraf{\pm 1}{l}{m}
\;\;.
\end{align}

Let us substitute expansion of type (\ref{49438148}) for electromagnetic quasibivectors $\bDB$ and $\bEH$ (\ref{YZ})
to equation (\ref{44396100}) without right-hand part.
Using (\ref{49414433}), (\ref{47086966}) and extracting coefficients for elementary spherical waves
$\spheraf{j}{l}{m}\, \e{-\bhim\,\omega\,x^0}$ and
 $\spheraf{j}{l}{m}\, \e{-\bhim\,\omega\,x^0}$,
we obtain a system of equations for quasi-hyperscalar radial functions $\Fourier{\bDB_{j}^{lm}}$ and $\Fourier{\bEH_{j}^{lm}}$.
Making simplifying transformations for this system and introducing new unknown functions
\begin{align}
\nonumber
 \Fourier{\bDB_{+}^{lm}} &\doteqdot \Fourier{\bDB_{1}^{lm}} + \Fourier{\bDB_{-1}^{lm}}
\;\;,\quad
&\Fourier{\bEH_{+}^{lm}} &\doteqdot \Fourier{\bEH_{1}^{lm}} + \Fourier{\bEH_{-1}^{lm}}
\;\;,
\\
\label{77019220}
\Fourier{\bDB_{-}^{lm}} &\doteqdot \Fourier{\bDB_{1}^{lm}} - \Fourier{\bDB_{-1}^{lm}}
\;\;,\quad
&\Fourier{\bEH_{-}^{lm}} &\doteqdot \Fourier{\bEH_{1}^{lm}} - \Fourier{\bEH_{-1}^{lm}}
\;\;,
\end{align}
we obtain the following system of equations:
\begin{subequations}
\label{77438462}
\begin{align}
\label{77445462}
(r^2\,\Fourier{\bDB_{0}^{lm}})_{;r} + \bhim\,r\,\sqrt{l\,(l+1)}\, \Fourier{\bDB_{+}^{lm}} &= 0
\;\;,\\
\label{77449168}
(r\,\Fourier{\bEH_{+}^{lm}})_{;r} - \bhim\,r\,\omega\,\Fourier{\bDB_{-}^{lm}}
- \bhim\,\sqrt{l\,(l+1)}\, \Fourier{\bEH_{0}^{lm}} &= 0
\;\;,\\
\label{78614383}
(r\,\Fourier{\bEH_{-}^{lm}})_{;r} - \bhim\,r\,\omega\,\Fourier{\bDB_{+}^{lm}}  &= 0
\;\;,\\
\label{78652228}
r\,\omega\,\Fourier{\bDB_{0}^{lm}} + \sqrt{l\,(l+1)}\,\Fourier{\bEH_{-}^{lm}}   &= 0
\;\;.
\end{align}
\end{subequations}
where $(...)_{;r}$ is the derivative with respect to radius $r$.

Let us find solutions to equation system (\ref{77438462}) for the case $\bDB = \bEH$.
After differentiation (\ref{77445462}) and
making necessary substitutions we obtain the following second-order equation for $\Fourier{\bDB_{0}^{lm}}$:
\begin{equation}
\label{65167798}
\left\{r^2\,\frac{\df^2}{\df r^2} + 2\,r\,\frac{\df}{\df r} + \left[r^2\,\omega^2 - l\,(l+1)\right]\right\}
\left(r\,\Fourier{\bDB_{0}^{lm}}\right) = 0
\;\;.
\end{equation}

If we have function $\Fourier{\bDB_{0}^{lm}}$ as a solution to equation (\ref{65167798}) then functions
$\Fourier{\bDB_{+}^{lm}}$ and
$\Fourier{\bDB_{-}^{lm}}$ are found directly from (\ref{77445462}) and (\ref{78652228}) ($\Fourier{\bDB_{j}^{lm}} = \Fourier{\bEH_{j}^{lm}}$)
 respectively.

At first let us consider the static case $\omega = 0$. Thus according to (\ref{78652228}) we have $\Fourier{\bDB_{-}^{lm}} = 0$.
For $l=0$ system (\ref{77438462}) has only one solution.
For $l\geqslant 1$ we take a solution which is finite as $r\to \infty$. As result we have for $\omega=0$:
\begin{equation}
\label{62698268}
\Fourier{\bDB_{0}^{lm}} = \frac{C^{lm}}{r^{2+l}}
\;\;,\quad
 \Fourier{\bDB_{+}^{lm}} = -\sqrt{\frac{l}{l+1}}\,\frac{C^{lm}\,\bhim}{r^{2+l}}
\;\;,\quad
 \Fourier{\bDB_{-}^{lm}} = 0
\;\;,
\end{equation}
where $C^{lm}$ are  quasi-hyperscalar constants, $l\geqslant 0$.

For the case $\omega\neq 0$ equation (\ref{65167798}) has the form of equation for so-called spherical Bessel functions.
These function are expressed by the Bessel functions of half-integer order (see. \cite{AbramStig1964}).
We will use the spherical Bessel function of the third kind $h^{(1)}_l$ and $h^{(2)}_l$ for the representation
of solutions to equations
(\ref{65167798}). These functions are expressed by the Hankel functions of half-integer order.
Because of relation $h^{(2)}_l(z) = (-1)^l\, h^{(1)}_l(-z)$ we can use only the functions $h^{(1)}_l$ but with
its argument
taking as positive as negative values. We will use only real
values of the argument.
Hyperimaginary unit $\bhim$  must be substituted for imaginary one in these functions.
Let us introduce the designation $\RadfunS{l}{}$ for spherical Bessel functions $h^{(1)}_l$. Hyperreal part of this function
$\bRe\RadfunS{l}{}$ is the spherical Bessel function of the first kind and hyperimaginary part $\bIm\RadfunS{l}{}$ is
the spherical Bessel function of the second kind.
Hyperconjugate function $\hconj{\RadfunS{l}{}}$ corresponds to the spherical Bessel function of the third kind $h^{(2)}_l$.

According to \cite{AbramStig1964} we have the following useful formulas:
\begin{equation}
\label{34388341}
\RadfunS{l}{} (z) = \frac{\e{\bhim\,z}}{\bhim^{l+1}\,z}\,\sum_{l^{\prime}=0}^{l}\frac{(l+l^{\prime})!}{(l-l^{\prime})!\,l^{\prime}!}\,\left(-2\,\bhim\,z\right)^{-l^{\prime}}
\;\;,
\end{equation}
\begin{equation}
\label{63424063}
\e{\bhim\,z\,\cos\vartheta} = \sum_{l=0}^{\infty}\left(2\,l+1\right)\bhim^{l}\,\left[\bRe\RadfunS{l}{}(z)\right]
\left[\spherzonf{0}{l}{0}(\cos\vartheta)\right]
\;\;.
\end{equation}

It is convenient to introduce also the following
\begin{Def}
\label{35215416}
The radial spherical functions are called the functions
\begin{equation}
\label{62152803}
\RadfunS{l}{k_r} \doteqdot k_r^{l+1}\,\RadfunS{l}{}(k_r\,r)
\;\;.
\end{equation}
where $\RadfunS{l}{}(z)$ is the first spherical Bessel function of the third kind
 with hyperimaginary unit instead of imaginary one.
\end{Def}

According to (\ref{34388341})  we have the following form of radial spherical functions
(\ref{62152803}) for $k_r = 0$:
\begin{equation}
\label{73739062}
\RadfunS{l}{0} = -\frac{(2\,l)!}{2^{l}\,l!}\,\frac{\bhim}{r^{l+1}}
\;\;.
\end{equation}

In view of (\ref{73739062}) the using of the functions $\RadfunS{l}{k_r}$ gives a possibility to represent solutions to equation
system (\ref{77438462}) for the cases
$\omega \neq 0$ and $\omega = 0$ with an united formula (see. below (\ref{45659877})).

Taking into account $\hconj{\RadfunS{l}{} (z)} = (-1)^{l}\,\RadfunS{l}{} (-z)$ and using (\ref{62152803}) we obtain
\begin{equation}
\label{34495966}
\hconj{\RadfunS{l}{k_r}} = -\RadfunS{l}{-k_r}
\;\;.
\end{equation}

\begin{subequations}
\label{50717508}
Using recurrent relations for the spherical Bessel functions \cite{AbramStig1964} we obtain the following relations for
the radial spherical functions:
\begin{align}
\label{47056535}
&\frac{\df}{\df r}\,\RadfunS{l}{k_r} = k_r^2\,\RadfunS{l-1}{k_r} - \frac{l+1}{r}\,\RadfunS{l}{k_r}
= -\RadfunS{l+1}{k_r} + \frac{l}{r}\,\RadfunS{l}{k_r}
\;\;,
\\
\label{48078488}
&k^2_r\,\RadfunS{l-1}{k_r} + \RadfunS{l+1}{k_r} = \frac{2\,l +1}{r}\,\RadfunS{l}{k_r}
\;\;.
\end{align}
\end{subequations}

\begin{subequations}
Using the introduced radial spherical functions we can write the following form of solution to equation (\ref{65167798}):
\label{45659877}
\begin{equation}
\label{47558654}
\Fourier{\bDB_{0}^{lm}} = \frac{C^{lm}_{k_r}}{r}\,\RadfunS{l}{k_r}
\;\;,
\end{equation}
where $C^{lm}_{k_r}$ are free hyperscalar constants, $k_r^2=\omega^2$.

We consider the case $\omega\neq 0$.
From (\ref{78652228}) we have that $\Fourier{\bDB_{0}^{00}}=0$ for $\omega\neq 0$ and so we consider the values $l\geqslant 1$
for zonal index.
We obtain the functions $\Fourier{\bDB_{+}^{lm}}$ and $\Fourier{\bDB_{-}^{lm}}$ from $\Fourier{\bDB_{0}^{lm}}$ directly with the help of
equations (\ref{77445462}) and (\ref{78652228}) respectively. Then we check that the obtained solution satisfies to equations
(\ref{77449168}) and (\ref{78614383}).
At last, using relations (\ref{77019220}) and (\ref{47056535}) we obtain
\begin{equation}
\label{48174835}
\Fourier{\bDB_{\pm 1}^{lm}} =
\frac{C^{lm}_{k_r}}{2}\,\sqrt{\frac{l}{l+1}}\left[
\frac{\bhim\,\omega^2}{l}\,\RadfunS{l-1}{k_r} - \left(\frac{\bhim}{r} {}\pm{} \frac{\omega}{l}\right)\RadfunS{l}{k_r}
\right]
\;\;,
\end{equation}
where $l\geqslant 1$, $-l\leqslant m \leqslant l$, $k_r = \pm \omega$.
\end{subequations}

For the case $\omega = 0$ (and, consequently, $k_r = 0$) solution (\ref{45659877}) coincides with the solution given by formula
(\ref{62698268}) for $l\geqslant 1$. The case $\omega = 0$, $l = 0$ also can be represented by formula (\ref{45659877}), if we take
by definition $\omega^2/l = 0$ for this case.
According to (\ref{73739062}) we can obtain a connection between the free constants
$C^{lm}_0$ and $C^{lm}$ which used in formulas
(\ref{45659877}) and (\ref{62698268}).
Thus formula (\ref{45659877}) can be used both for static and time-periodical cases.
It give the solutions which decrease at $r$-infinity.

Let us introduce the following designation:
\begin{equation}
\label{46736430}
\garmsb{\omega}{\pm}{l}{m} \doteqdot \Fourier{\bDB_{0}^{l}}\,\spheraf{0}{l}{m} + \Fourier{\bDB_{1}^{l}}\,\spheraf{1}{l}{m}
+ \Fourier{\bDB_{-1}^{l}}\,\spheraf{-1}{l}{m}
\;\;,
\end{equation}
where $\Fourier{\bDB_{j}^{l}}\doteqdot \Fourier{\bDB_{j}^{lm}}$ from (\ref{45659877}) with substitution $C^{lm}_{k_r} = 1$,
left-hand index $+$ or $-$ corresponds to one case from the two $k_r = \pm\omega$.

Using (\ref{55376750}) and (\ref{34495966})-(\ref{46736430}) we obtain the rule for hyperconjugation
\begin{equation}
\label{34669816}
\hconj{\garmsb{\omega}{\pm}{l}{m}} = (-1)^{m+1}\,\garmsb{\omega}{\mp}{l}{,-m}
\;\;.
\end{equation}

Thus an elementary spherical solution to equation (\ref{44396100}) without sources and for the case $\bDB=\bEH$
can be written in the form
\begin{equation}
\label{76567890}
\garmsb{\omega}{\pm}{l}{m}\, \e{-\bhim\, \omega\,x^0}
\;\;.
\end{equation}

It is evident that $\garmsb{\omega}{\pm}{l}{m}$ are eigenfunctions of self-adjoint operator
$-\bhim\,\bnp$:
\begin{equation}
\label{47861464}
-\bhim\,\bnp\,\garmsb{\omega}{\pm}{l}{m} = \omega\,\garmsb{\omega}{\pm}{l}{m}
\;\;.
\end{equation}

Using (\ref{34388341}) we can conclude that
the function $\garmsb{\omega}{+}{l}{m}\,\e{-\bhim\,\omega\,x^0}$ is a radially divergent elementary spherical wave and
$\garmsb{\omega}{-}{l}{m}\,\e{-\bhim\,\omega\,x^0}$ is a convergent one. The sign of circular frequency $\omega$ influences to
its polarization but not to the direction of propagation.

In general case, for given time harmonic $\e{-\bhim\,\omega\,x^0}$ (harmonics differing with the sign of the cyclic frequency $\omega$
is considered as different) we must take the sum of divergent and convergent waves with arbitrary quasi-hyperscalar coefficients.
We have an everywhere regular solution if we take hyperreal parts of radial spherical functions
$\bRe\RadfunS{l}{k_r}$ in expressions (\ref{45659877}).
Thus in view of (\ref{34495966}) we can write the following everywhere regular and finite functions:
\begin{equation}
\label{60230079}
\garmsb{\omega}{\bumpeq}{l}{m} \doteqdot \frac{1}{2}\left(\garmsb{\omega}{+}{l}{m} - \garmsb{\omega}{-}{l}{m}\right)
\;\;.
\end{equation}
The appropriate elementary solutions
\begin{equation}
\label{48141318}
\garmsb{\omega}{\bumpeq}{l}{m}\,\e{-\bhim\,\omega\,x^0}
\end{equation}
are regular standing waves with descending amplitude as $r\to\infty$.

Any regular solution to equation (\ref{44396100}) for the case $\bDB=\bEH$ and $\bjem=0$ can be represented as
a sum (and integral with respect to $\omega$ from $-\infty$ to $\infty$) of elementary spherical waves (\ref{48141318})
with some hyperscalar coefficients.

Let us find this representation for plane waves $\cylaf{1}{1}\,\e{\bhim\,\omega\,(x^3 - x^0)}$ (\ref{55859098}) and
$\cylaf{-1}{-1}\,\e{-\bhim\,\omega\,(x^3 + x^0)}$ (\ref{55867560}).
Obviously, it will suffice to obtain the expansion of bivector function $\cylaf{1}{1}\,\e{\bhim\,\omega\,x^3}$ in term of
bivector functions $\garmsb{\omega}{\bumpeq}{l}{m}$ (\ref{60230079}). For the mode propagating in the negative direction of $x^3$ axis
we can use the hyperconjugation operation $\hconj{(\cylaf{1}{1}\,\e{\bhim\,\omega\,x^3})} = \cylaf{-1}{-1}\,\e{-\bhim\,\omega\,x^3}$.

Using (\ref{63424063}) and (\ref{62152803}) we can write
\begin{equation}
\label{43415984}
\cylaf{1}{1}\,\e{\bhim\,\omega\,x^3} =
\cylaf{1}{1}\,\e{\bhim\,\omega\,r\,\cos\vartheta}
= \sum_{l=0}^{\infty}\left(2\,l+1\right)\bhim^{l}\,\omega^{-l-1}\,
\spherzonf{0}{l}{0}\,\cylaf{1}{1}\,\bRe\RadfunS{l}{\omega}
\;\;.
\end{equation}

To expand $\spherzonf{0}{l}{0}\,\cylaf{1}{1}$ let us use the multiplication tables which we have.
Converting relation (\ref{62941079}) and using (\ref{47394428}) we have, in particular, the following
relation:
\begin{equation}
\label{44625493}
\cylaf{1}{1} = \spheraf{1}{1}{1} + \spheraf{-1}{1}{1} - \bhim\,\sqrt{2}\,\spheraf{0}{1}{1}
\;\;.
\end{equation}
According to (\ref{45992823}) and (\ref{46589541}) we can obtain
\begin{subequations}
\label{45281065}
\begin{align}
\label{45290048}
&\spherzonf{0}{l}{0}\,\spherzonf{0}{1}{1} =
\frac{1}{\sqrt{2}\,(2\,l+1)}\left[-\sqrt{l\,(l-1)}\;\spherzonf{0}{l-1}{1} + \sqrt{(l+1)\,(l+2)}\;\spherzonf{0}{l+1}{1}\right]
\;,\quad\\
&\spherzonf{0}{l}{0}\,\spherzonf{\pm 1}{1}{,1} =
\frac{1}{2}\left\{
\pm\spherzonf{\pm 1}{l}{,1}
\label{45292171}
+ \frac{1}{2\,l+1}\left[(l-1)\,\spherzonf{\pm 1}{l-1}{,1}  + (l+2)\,\spherzonf{\pm 1}{l+1}{,1}\right]
\right\}
\;,
\end{align}
\end{subequations}
where  $l\geqslant 1$.

Using (\ref{44625493}), (\ref{45281065}), and (\ref{47394428}) we have
\begin{multline}
\label{46312378}
\spherzonf{0}{l}{0}\,\cylaf{1}{1} =
\frac{1}{2} \left(\spheraf{1}{l}{1} - \spheraf{-1}{l}{1}\right)
\\
+ \frac{1}{2\,(2\,l+1)}\left[(l-1)\left(\spheraf{1}{l-1}{,1} + \spheraf{-1}{l-1}{,1}
\right) + (l+2)\left(\spheraf{1}{l+1}{,1} + \spheraf{-1}{l+1}{,1}\right)\right]
\\
+ \frac{\bhim}{2\,l+1}\left[\sqrt{l\,(l-1)}\;\spheraf{0}{l-1}{,1} - \sqrt{(l+1)\,(l+2)}\;\spheraf{0}{l+1}{,1}\right]
\;\;.
\end{multline}

Substituting (\ref{46312378}) into (\ref{43415984}), grouping the terms with identical $l$ indices, using (\ref{48078488}), and
comparing with (\ref{45659877}), (\ref{46736430}), (\ref{60230079}) we obtain the following expansion:
\begin{equation}
\label{44653138}
\cylaf{1}{1}\,\e{\bhim\,\omega\,r\,\cos\vartheta} =
-\sum_{l=1}^{\infty}
\bhim^{l}\,\sqrt{l\,(l+1)}\;(2\,l+1)\,\omega^{-l-2}\garmsb{\omega}{\bumpeq}{l}{1}
\;\;.
\end{equation}

\section{Conclusions}
\label{concl}
Thus the basic systems of orthogonal functions for space-time multivectors are built at the present work.
This work considers, in particular, cylindrical and spherical space-time multivector functions with special emphasis
on their application to nonlinear electrodynamics. Obtained appropriate systems of first order differential equations
(\ref{42346412}) and (\ref{77438462}) can be used for various problems of nonlinear and linear electrodynamics.
Non-commutative multiplication tables for cylindrical and spherical bivector functions obtained in this work
are of considerable importance in nonlinear and linear (with non-homogeneous medium) problems.
Also these multiplication tables are very helpful in symbolic computing for the mathematical problems.

\appendix

\renewcommand{\theequation}{\Alph{section}.\arabic{equation}}

\section*{Main designations}
\addcontentsline{toc}{section}{Main designations}
\label{REPROTdst}
\begin{tabular*}{\textwidth}{|c|l@{\extracolsep{\fill}\hspace*{2mm}}|c|}
\hline
Symbol & Name & Appearance\\
\hline
\hline $\baab^\mu$, $\bbaab^i$ & \parbox[t]{40ex}{Basis vectors and bivectors} & (\ref{54511673})\\
\hline $\unitc$; $\bhim$, $\him^{\mu\nu\rho\sigma}$ & \parbox[t]{40ex}{Hyperunit and hyperimaginary unit} & (\ref{54511673}),(\ref{56226711})\\
\hline $\hconj{\mathbf{C}}$ & \parbox[t]{40ex}{Hyperconjugation} & (\ref{55221747})\\
\hline $\bRe\mathbf{C}$, $\bIm\mathbf{C}$ & \parbox[t]{40ex}{Hyperreal and hyperimaginary parts} & (\ref{40209010})\\
\hline $\metr_{\mu\nu}$ & \parbox[t]{40ex}{Metric tensor} & (\ref{56226711})\\
\hline $\bcdot$, $\bwedge$ & \parbox[t]{40ex}{Symmetrical and asymmetrical products} & (\ref{64298295})\\
\hline $\bFem$, $\Fem_{\mu\nu}$; $\bfem$, $\fem^{\mu\nu}$ & \parbox[t]{40ex}{Bivectors of electromagnetic field} & (\ref{HyperEMF})\\
\hline $\bEem$, $\Eem_i$; $\bHem$, $\Hem_i$ & \parbox[t]{40ex}{Intensities of  fields:\\ electric and magnetic} & (\ref{HyperEMF})\\
\hline $\bDem$, $\Dem^i$; $\bBem$, $\Bem^i$ & \parbox[t]{40ex}{Inductions: electric and magnetic} & (\ref{HyperEMF})\\
\hline $\bnp$; $\bnpc$ & \parbox[t]{40ex}{Operator of space differentiation,\\ its angle part} & (\ref{defbp}),(\ref{49414433})\\
\hline $\bDB$; $\bEH$ & \parbox[t]{40ex}{Quasi-bivectors of electromagnetic field} & (\ref{YZ})\\
\hline $\prodf{\st{1}{\mathbf{C}}}{\st{2}{\mathbf{C}}}$ & \parbox[t]{40ex}{Functional product} & (\ref{kkklikdjjff})\\
\hline $\norma{\mathbf{C}}$ & \parbox[t]{40ex}{Norm of functional vector} & (\ref{41825843})\\
\hline $\conjop{\mathbf{Q}}$ & \parbox[t]{40ex}{Adjoint operator} & (\ref{ddjekks})\\
\hline $\bShST_\mu$; $\bRotS_i$ & \parbox[t]{40ex}{Self-adjoint infinitesimal operators\\ for shift and rotation} & (\ref{sskkeiir}),(\ref{kkjnmfndd})\\
\hline $\commut{\mathbf{A}}{\mathbf{B}}$ & \parbox[t]{40ex}{Commutator for operators} & (\ref{39747725})\\
\hline $\Fourier{\mathbf{Q}}$ & \parbox[t]{40ex}{Fourier transform or coefficient} & (\ref{wwqsder})\\
\hline $\cylaf{j}{m}$ & \parbox[t]{40ex}{Angle cylindrical functions} & (\ref{44146270})\\
\hline $\RadfunC{|m|}{k_\rho}$; $\RadfunS{l}{k_r}$ & \parbox[t]{40ex}{Radial functions:\\ cylindrical and spherical} & (\ref{75002577}),(\ref{62152803})\\
\hline $\garmcb{\omega}{k_3}{\pm}{m}$; $\garmsb{\omega}{\pm}{l}{m}$ & \parbox[t]{40ex}{Cylindrical and spherical\\ bivector eigenfunctions of operator $-\bhim\,\bnp$} & (\ref{53130640}),(\ref{46736430})\\
\hline $\spheraf{j}{l}{m}$; $\sphersecf{j}{m}$; $\spherzonf{j}{l}{m}$, $\zc$ & \parbox[t]{40ex}{Angle spherical functions:\\ general, sectorial, zonal, its argument} & (\ref{48706537}),(\ref{47394428})\\
\hline $\nosum$ & \parbox[t]{40ex}{Cancellation of summation\\ on repeating indices} & (\ref{47394428})\\
\hline $\ClGord{l_1}{l_2}{l}{m_1}{m_2}$ & \parbox[t]{40ex}{Clebsch-Gordan coefficients\\ for rotation group} & (\ref{45992823})\\
\hline
\end{tabular*}

\end{document}